\documentclass[lettersize,journal]{IEEEtran} 
\usepackage{changepage}
\usepackage{cite}
\usepackage{amsmath,amssymb,amsfonts}
\usepackage{algpseudocode}
\usepackage{textcomp}
\usepackage{xcolor}
\usepackage{amsmath}
\usepackage{tabularx}
\usepackage{amsthm}
\usepackage{multirow}
\usepackage{relsize}
\usepackage{optidef}
\usepackage{bbold}
\usepackage{graphicx,subcaption,caption}
\usepackage{import}
\usepackage{array} 
\usepackage{bm, cuted}
\usepackage{tikz}
\usepackage{mathtools}
\usepackage{bbm}
\usepackage{cite}
\usepackage{algorithm}
\usepackage{algpseudocode}
\usetikzlibrary{positioning, arrows.meta, calc, intersections, shapes, decorations.pathreplacing}
\usepackage{upgreek}
\usepackage{comment}
\usepackage{bm}
\usepackage{fancyhdr}
\newcommand{\E}{\mathbb{E}}

\newcommand{\argmin}{\operatorname*{argmin}}

\algrenewcommand\algorithmicrequire{\textbf{Input:}}
\algrenewcommand\algorithmicensure{\textbf{Output:}}

\makeatletter
\newcommand{\removelatexerror}{\let\@latex@error\@gobble}
\makeatother

\newcommand{\FigFont}{\fontsize{9}{10.5}\selectfont}
\newcommand{\FigTitleFont}{\fontsize{9.5}{11}\selectfont\bfseries}
\newcommand{\FigSmallFont}{\fontsize{8}{9.5}\selectfont}

\tikzset{
    pics/basestation/.style={
        code={
            \begin{scope}[scale=0.3]
                \definecolor{towerblue}{rgb}{0.75,0.88,1}

                \begin{scope}[shift={(0,5.6)}]
                  \fill[towerblue] (-1,0) rectangle (1,1.2);
                  \foreach \x in {-0.75,-0.5,...,0.75}
                    \foreach \y in {0.15,0.45,...,1.05}
                      \fill[white] (\x,\y) circle (1.5pt);
                \end{scope}

                \path[name path=tower,blue!60] (-1.6,0) -- (-0.6,5) -- (0.6,5) -- (1.6,0) -- cycle;
                \fill[towerblue] (-1.6,0) -- (-0.6,5) -- (0.6,5) -- (1.6,0) -- cycle;
                \draw[thick] (-1.6,0) -- (-0.6,5) -- (0.6,5) -- (1.6,0) -- cycle;

                \begin{scope}
                  \clip (-1.6,0) -- (-0.6,5) -- (0.6,5) -- (1.6,0) -- cycle;
                  \foreach \y in {0.4, 1.2, 2.0, 2.8, 3.6, 4.4} {
                    \draw[blue!60, line width=3pt] (-1.6,\y-0.4) -- (1.6,\y+0.4);
                    \draw[blue!60, line width=3pt] (1.6,\y-0.4) -- (-1.6,\y+0.4);
                  }
                \end{scope}

                \draw[blue!60, line width=4pt] (0,5) -- (0,5.6);
            \end{scope}
        }
    }
}

\tikzset{
    pics/ue/.style={
        code={
            \begin{scope}[scale=0.25, line width=0.8pt]
                \shade[rounded corners=2pt, left color=blue!10, right color=blue!30]
                  (0,0) rectangle (2,4);
                \draw[rounded corners=2pt] (0,0) rectangle (2,4);

                \shade[rounded corners=1pt, inner color=gray!10, outer color=blue!20]
                  (0.4,1.2) rectangle (1.6,3.4);
                \draw[black, thick, rounded corners=2pt] (0.4,1.2) rectangle (1.6,3.4);

                \shade[ball color=blue!30!white] (1,0.5) circle (0.3);
                \draw[black, thick] (1,0.5) circle (0.3);

                \draw[rounded corners=2pt, fill=black!70] (0.9,4) rectangle (1.1,4.4);

                \foreach \r in {0.4, 0.6, 0.8} {
                  \draw[blue!60, thick] (1-\r,4.4+\r) arc[start angle=180,end angle=0,radius=\r];
                }
            \end{scope}
        }
    }
}

\tikzset{
    pics/syncTx/.style={
        code={
            \begin{scope}[font=\FigFont, every node/.style={align=center}]
                \node[font=\FigTitleFont] at (0, 1.2) {Synchronous Transmission};
                
                \node[draw=black, rounded corners=6pt, thick, fill=green!15, inner sep=10pt] (eqbox)
                {$
                    x_k^{\rm d}(t) = \sum \limits_{i=1}^{N_\text{u}} \sqrt{p_k} b_k[i] p(t - (i-1)T)
                $};
            \end{scope}
        }
    }
}

\tikzset{
    pics/asyncTx/.style={
        code={
            \begin{scope}[font=\FigFont, every node/.style={align=center}]
                \node[font=\FigTitleFont] at (0, 1.2) {Asynchronous Transmission};
                
                \node[draw=black, rounded corners=6pt, thick, fill=green!15, inner sep=10pt] (eqbox)
                {$
                    x_k^{\rm d}(t) = \sum \limits_{i=1}^{N_\text{u}} \sqrt{p_k} b_k[i] p(t - (i-1)T - \tau_kT)
                $};
            \end{scope}
        }
    }
}

\tikzset{
    pics/asyncRx/.style={
        code={
            \begin{scope}[font=\FigFont, every node/.style={align=center}]
                \node[font=\FigTitleFont] at (-1, 2.95) {Asynchronous Reception};

                \draw[rounded corners=6pt, line width=1.4pt, fill=yellow!20]
                  (-6,-2.35) rectangle (4,2.45);

                \draw[rounded corners=4pt, line width=1.4pt, fill=gray!10]
                  (-5.5,-0.5) rectangle (-2,0.5);
                \node at (-3.75,0) {$\boldsymbol{y}_\ell(t) * p(t)$};

                \coordinate (A) at (-2,0);
                \coordinate (F) at (-1,0);
                \draw[line width=1.4pt] (A) -- (F);

                \coordinate (B1) at (0,1.10);
                \coordinate (B3) at (0,-1.10);

                \coordinate (C1) at (1.2,1.10);
                \coordinate (C3) at (1.2,-1.10);

                \coordinate (D1) at (0.05,1.55);
                \coordinate (D3) at (0.05,-0.55);

                \draw[line width=1.4pt] (F) |- (B1);
                \draw[line width=1.4pt] (F) |- (B3);

                \fill[blue!60, draw=black, line width=1pt] (B1) circle (3pt);
                \fill[blue!60, draw=black, line width=1pt] (C1) circle (3pt);
                \draw[line width=1.4pt] (D1) -- (C1);
                \draw[line width=1.4pt] (C1) -- ($(C1)+(1,0)$);

                \fill[blue!60, draw=black, line width=1pt] (B3) circle (3pt);
                \fill[blue!60, draw=black, line width=1pt] (C3) circle (3pt);
                \draw[line width=1.4pt] (D3) -- (C3);
                \draw[line width=1.4pt] (C3) -- ($(C3)+(1,0)$);

                \node at (1,0) {$\vdots$};
                \node at (3,0) {$\vdots$};

                \node[font=\FigSmallFont, anchor=south] at (1.00,1.62)
                {$t=(j-1)T+\tau_1T$};

                \node[font=\FigSmallFont, anchor=north] at (1.00,-1.42)
                {$t=(j-1)T+\tau_KT$};

                \node[anchor=west] at ($(C1)+(1.1,0)$) {$\boldsymbol{y}_{\ell,1}[j]$};
                \node[anchor=west] at ($(C3)+(1.1,0)$) {$\boldsymbol{y}_{\ell,K}[j]$};
            \end{scope}
        }
    }
}

\tikzset{
    pics/syncRx/.style={
        code={
            \begin{scope}[font=\FigFont, every node/.style={align=center}]
                \node[font=\FigTitleFont] at (0, 1.75) {Synchronous Reception};

                \draw[rounded corners=6pt, line width=1.4pt, fill=yellow!20]
                  (-5,-1.10) rectangle (5,1.20);

                \draw[rounded corners=4pt, line width=1.4pt, fill=gray!10]
                  (-4.5,-0.5) rectangle (-1,0.5);
                \node at (-2.75,0) {$\boldsymbol{y}_\ell(t) * p(t)$};

                \coordinate (A) at (-1,0);
                \coordinate (B) at (0,0);
                \draw[line width=1.4pt] (A) -- (B);

                \coordinate (C) at (1.2,0);
                \coordinate (D) at (0.05,0.45);

                \fill[blue!60, draw=black, line width=1pt] (B) circle (3pt);

                \fill[blue!60, draw=black, line width=1pt] (C) circle (3pt);

                \draw[line width=1.4pt] (D) -- (C);

                \draw[line width=1.4pt] (C) -- ($(C)+(1,0)$);

                \node[font=\FigSmallFont, anchor=south] at (0.78,0.40)
                {$t=(j-1)T$};

                \node[anchor=west] at ($(C)+(1.3,0)$) {$\boldsymbol{y}_\ell[j]$};
            \end{scope}
        }
    }
}

\fancypagestyle{firstpage}{%
  \fancyhf{} 
  \fancyhead[C]{%
    \begin{minipage}{\textwidth}
    \centering
    \scriptsize
    This work has been submitted to the IEEE for possible publication. 
    Copyright may be transferred without notice, after which this version may no longer be accessible.
    \end{minipage}%
  }%
}

\begin{document}
\title{Asynchronous Cell-Free Wireless Networks}

\author{Hossein Maleki$^{1}$, \IEEEmembership{Graduate Student Member, IEEE}, Angel Lozano$^{2}$, \IEEEmembership{Fellow, IEEE}, and Hamid Jafarkhani$^{1}, \IEEEmembership{Fellow, IEEE}$
\thanks{$^{1}$ H. Maleki and H. Jafarkhani are with the Department of Electrical Engineering and Computer Science, Univ. of California, Irvine. Their work was supported in part by NSF Award CCF-2328075. Emails:         {\tt\small \{malekih, hamidj\}@uci.edu }} %
\thanks{$^{2}$ A. Lozano is with Universitat Pompeu Fabra (UPF), Barcelona. His work was supported by the Maria de Maeztu Units of Excellence Programme CEX2021-001195-M funded by MICIU/AEI/10.13039/501100011033, and by AGAUR. Email:        {\tt\small  angel.lozano@upf.edu }}
}

\maketitle
\thispagestyle{firstpage}
\begin{abstract}

This paper introduces intentional symbol-level asynchronicity as an interference-management mechanism for the uplink of cell-free networks. Rather than treating timing offsets as impairments, user-specific delays are deliberately assigned to reduce multiuser interference and improve spectral efficiency. A scalable receiver architecture is devised with synchronous channel-estimation pilots, asynchronous data transmission, bounded matched-filter sampling, and windowed combining. The delay assignment is formulated by means of a conflict metric that measures the severity of multiuser interference and is solved through a low-complexity greedy algorithm. As a benchmark, an exact method is also formulated for small and moderate network sizes using mixed-integer linear programming. Numerical results with both rectangular and root-raised-cosine pulses confirm that considerable performance improvements are feasible relative to synchronous transmissions, even with
small delay codebooks and short observation windows.
\end{abstract}

\begin{IEEEkeywords}
Cell-free networks, asynchronous uplink, delay assignment, matched filtering, user-centric processing, mixed-integer linear programming, scalability
\end{IEEEkeywords}

\IEEEpeerreviewmaketitle

\section{Introduction}

In a cell-free (CF) network, many geographically distributed access points (APs) jointly serve user equipments (UEs) over common time/frequency resources.
Such CF networks can provide a higher coverage probability and a more uniform quality of service than traditional cellular deployments \cite{ngo2017cellfree,buzzi2017usercentric}. These advantages, though, come at the expense of new challenges, including channel estimation, combining, precoding, deployment, and AP-UE association rules that remain scalable as the number of APs and UEs grows \cite{9930941,10186347}. The scalability issue in particular has motivated CF propositions based on subsets, clusters, and variants thereof \cite{9043895,bjornson2020competitive,bjornson2020scalable,buzzi2020usercentric,zhang2021local,shaik2021mmse,munawar2024architecture}.

Early studies established the fundamentals of precoding, power control, spectral efficiency, and energy efficiency \cite{nayebi2017precoding,ngo2018total}, with fairness rules and high-resolution quantization shown to be of central importance \cite{bashar2019uplink,nguyen2020maxmin,bashar2019energy}. A parallel line of work characterized practical constraints such as fronthaul constraints, \cite{10186347,masoumi2020performance,demirhan2022enabling}, channel estimation \cite{interdonato2019downlink,jin2019channel}, multi-antenna UEs \cite{wang2022uplink}, and resource allocation \cite{ammar2022distributed}, sometimes applying stochastic geometry \cite{papazafeiropoulos2020performance}. More recently, CF architectures have been integrated with reconfigurable intelligent surfaces to further improve coverage and efficiency \cite{vanchien2022reconfigurable,shi2022spatially,gan2022multiple,nguyen2023spectral} and are used in applications in which access to cellular infrastructure does not exist or is damaged~\cite{10834526}.

Most CF formulations assume that the uplink signals reach the distributed receivers synchronously, yet this assumption is questionable because APs and UEs are separated by different distances, have independent oscillators, and may experience timing offsets. Even with perfect synchronicity at a given node, there would be no guarantee of synchronicity elsewhere. Thus, asynchronicity is perceived as a potential impairment and has been studied as such. In \cite{yan2019asynchronousDMmimo}, it was shown that the phase shifts caused by timing offsets substantially reduce the signal power. The study in \cite{li2021impacts} revealed that asynchronous reception degrades channel estimates, especially at high signal-to-noise ratio. In turn, \cite{zheng2022performance,zheng2023asynchronousRS} analyzed asynchronous reception under propagation delay differences and oscillator-induced phase shifts; these effects were shown to disrupt pilot orthogonality and coherent data transmission. For its part, \cite{li2025mixed} studied asynchronous orthogonal frequency-division multiplexing (OFDM) with mixed coherent and non-coherent transmissions, suggesting the clustering of APs into coherent groups. Then, \cite{xin2024pbta} studied asynchronous reception in wide-area scenarios, where timing offsets can exceed the cyclic prefix duration; it proposed a per-beam timing advance to align signals in the beam domain and suppress asynchronous interference.
Channel estimation was also the subject of \cite{sun2025channelestimation}, which proposed extended pilot sequences to recover pilot orthogonality.

In contrast with all the foregoing works, which regarded asynchronicity as
a detrimental phenomenon that should be mitigated, this work embraces asynchronicity. The goal is to formulate CF networks that can harness
\emph{intentional} timing offsets that, rather than being at the mercy of the propagation and the hardware, are carefully designed.
This is motivated by fundamental results supporting how asynchronous signaling can enlarge the capacity region of multiple-access channels  \cite{verdu1989asynchronous,8618621,ganji2021asynchronous,10720669}. 
Underlying this improvement is the notion of sampling diversity, whereby the redundancy of the asynchronous waveforms can be exploited \cite{ganji2016samplingdiversity}. 
For the two-user case specifically,
the sum-capacity depends on \cite{ganji2021asynchronous} 
\begin{equation}\label{eq:async_mac_sum_rate_term}
    1+\frac{S_1(f)}{\sigma_1^2} + \frac{S_2(f)}{\sigma_2^2} +\frac{S_1(f)S_2(f)}{\sigma_1^2\sigma_2^2} \left(1-G^2(f)\right),
\end{equation}
where $S_k(f)$ is the power spectral density of User $k$, $\sigma_k^2$ is its noise level, and $G(f)$ the folded spectrum that hinges on the pulse shape and delay.
For synchronous transmissions with a Nyquist pulse, $G(f)=1$ and hence the last term in \eqref{eq:async_mac_sum_rate_term} vanishes. With asynchronous transmissions, though, $G(f)\leq 1$ and thus the rate is enhanced by $\big( 1-G^2(f) \big)$.


This result on the multiple-access channel, which has
demonstrated 
that timing mismatches provide an additional user separability mechanism besides time, frequency, code, or space \cite{das2011mimo,ganji2016samplingdiversity,ganji2021asynchronous,maleki2026delay},
is the starting point for the present work. The goal is to determine
whether controlled timing offsets can serve as a scalable
interference-management instrument in CF networks.
The paper tackles this question by developing a conflict-aware asynchronous CF uplink architecture.
Timing offsets drawn from a codebook are assigned, such that UEs that are highly nonseparable in the spatial domain experience delays 
that reduce pulse-overlap interference. What is sought is not to render all the transmissions arbitrarily asynchronous. Rather, the asynchronicity is carefully controlled and each AP needs to sample only the delay branches associated with its served users. This ensures that the local sampling and detection dimensions remain bounded as the network grows large, which is crucial to preserve the scalability.

The main contributions are:
\begin{itemize}
    \item A matched-filter (MF) uplink signal model for CF networks, with synchronous pilots and controlled asynchronous data transmissions. The model accounts for imperfect channel estimation, residual pilot leakage, and colored noise induced by fractional-delay sampling.
    \item A codebook sampling architecture in which each AP samples only a bounded set of delay branches, determined by its served UEs. This reduces the sampling burden, and the overall computational complexity.
    \item A pairwise interference surrogate based on pulse-overlap energy.
    This surrogate is then related to a lower bound on the signal-to-interference-plus-noise ratio (SINR) attained with minimum mean-square error (MMSE) reception.
    \item 
    A windowed MMSE combiner that employs only the MF samples near the target symbol. As distant samples contribute only weakly, this structure approaches the performance of full-frame MMSE reception with a much smaller combiner dimension.
    \item A formulation of the resulting delay-assignment problem over a finite delay codebook, presented as an exact mixed-integer linear program (MILP), along with a greedy weighted-coloring algorithm to solve it. The MILP allows gauging, for small and moderate network sizes, the optimality of scalable heuristics.
\end{itemize}

The remaining sections proceed as follows.
Sec.~\ref{sec:sys_model} introduces the system and signal models. Then, Sec.~\ref{sec:uplink_reception} devises channel estimation, as well as our centralized and local reception schemes with synchronous pilots and asynchronous data. Sec.~\ref{sec:interference_surrogate} formulates the interference-based delay-assignment metric, while Sec.~\ref{sec:delay_assignment} presents our exact and greedy assignment methods to solve it. Sec.~\ref{sec:simulation_results} evaluates the proposed solutions, and Sec.~\ref{sec:conclusion} concludes the paper.

\section{System Model}\label{sec:sys_model}

The system under consideration features $L$ APs and $K$ UEs, with each AP having $N_\text{R}$ antennas and each UE having a single antenna. The focus is on the uplink.

\subsection{Channel Model}

The channel vector from the $k$th UE and the $\ell$th AP is
\begin{align}
\label{eq:channel}
    \boldsymbol{h}_{\ell, k}\sim \mathcal{N}_{\mathbb{C}}(\boldsymbol{0},\boldsymbol{R}_{\ell, k}),
\end{align}
where
\begin{align}
    \label{eq:channel_cov}
\boldsymbol{R}_{\ell, k}=\beta_{\ell, k}\boldsymbol{\Theta}_{\ell, k}
\end{align}
is the spatial covariance matrix, with
$\operatorname{tr}(\boldsymbol{\Theta}_{\ell, k})=N_\text{R}$
and with the large-scale channel gain
\begin{align} \label{eq:channel_power}
    \beta_{\ell, k} = \frac{\beta_0}{D_{\ell, k}^{\kappa}},
\end{align}
where $\beta_0$ and $\kappa$ denote the gain at a reference distance of $1$~m and the pathloss exponent, respectively, and $D_{\ell, k}$ is the distance.
The channels are block-faded with $N_\text{c}$ symbols per coherence block, of which $N_\text{p}$ are reserved for pilot transmission and the remainder $N_\text{d} = N_\text{c} - N_\text{p}$ are for payload data.

\subsection{Signal Model}

\begin{figure}[]
     \centering
     \begin{tikzpicture}[scale=1.2]

\definecolor{lightblue}{RGB}{200,220,255}
\definecolor{lightorange}{RGB}{255,210,160}
\definecolor{lightgreen}{RGB}{200,255,200}

\def\recth{0.7}
\def\rectw{2}

\def\twoffset{0.5}
\def\throffset{1}

\def\ya{0}
\def\yb{-1.5}
\def\yc{-3}

\foreach \y in {\ya, \yb, \yc} {
    \draw[dashed] (-1.5,\y) -- (5,\y);
}

\fill[lightblue] (0,\ya) rectangle (\rectw,\ya+\recth);
\fill[lightorange] (\twoffset,\yb) rectangle (\twoffset+\rectw,\yb+\recth);
\fill[lightgreen] (\throffset,\yc) rectangle (\throffset+\rectw,\yc+\recth);

\node at (1,\ya+0.35) {\small $s_1[i]$};
\node at (\twoffset+1,\yb+0.35) {\small $s_2[i]$};
\node at (\throffset+1,\yc+0.35) {\small $s_3[i]$};

\node[left] at (-0.9,\ya+0.35) {\small $s_1[i-1]$};
\node[left] at (-0.4,\yb+0.35) {\small $s_2[i-1]$};
\node[left] at (+0.05,\yc+0.35) {\small $s_3[i-1]$};

\node[right] at (2,\ya+0.35) {\small $s_1[i+1]$};
\node[right] at (2.8,\yb+0.35) {\small $s_2[i+1]$};
\node[right] at (3.2,\yc+0.35) {\small $s_3[i+1]$};

\draw[very thick, blue] (0,\yc-0.2) -- (0,\ya+\recth+0.2);

\node[above left] at (0,\ya+\recth+0.2) {\scriptsize $\tau_1T=0$};

\draw[|-|, thick] (0,\yb+0.35) -- (\twoffset,\yb+0.35);
\node[above] at (\twoffset/2,\yb+0.5) {\scriptsize $\tau_2T$};

\draw[|-|, thick] (0,\yc+0.35) -- (\throffset,\yc+0.35);
\node[above] at (\throffset/2,\yc+0.5) {\scriptsize $\tau_3T$};

\draw[<->, thick] (0,\ya+\recth+0.6) -- (\rectw,\ya+\recth+0.6);
\node at (1,\ya+\recth+0.85) {$T$};

\end{tikzpicture}
     \caption{Asynchronous transmission for $K=3$ UEs with rectangular pulses.}
     \label{fig:f1}
\end{figure}
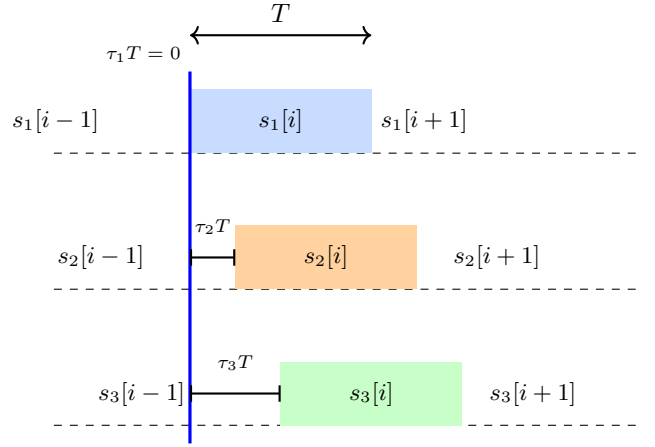

To preserve pilot orthogonality and facilitate channel estimation, pilot transmissions remain synchronous.
Thus, letting $\boldsymbol{\phi}_k = [\phi_k[-(N_\text{p}-1)], \dots, \phi_k[0]]^\top$ be a vector containing the pilot sequence of User $k$, 
\begin{align}\label{eq:orthogonal_pilots}
    \boldsymbol{\phi}_k^*\boldsymbol{\phi}_m =
    \begin{cases}
        N_\text{p} & k=m\\
        0 & k\neq m.
    \end{cases}
\end{align}
Subsequently, User $k$ proceeds to transmit the payload data sequence
$\boldsymbol{s}_k = [s_k[1], \dots, s_k[N_\text{d}]]^\top$, satisfying $\mathbb{E}[ \boldsymbol{s}_k \boldsymbol{s}_k^*]= \boldsymbol{I}_{N_\text{d}}$.
User $k$'s data are subject to a delay $\tau_k T$, where $T$ is the symbol period, $\tau_k \in [0,1)$. This is illustrated in Fig.~\ref{fig:f1} for $K=3$ where without loss of generality, we assumed $\tau_1 = 0$.  Then, at time $t$, the $k$th UE transmits
\begin{align} \label{eq:tx_sig_syncpilot_asyncdata}
    x_k(t) &= \sum_{i=-(N_\text{p}-1)}^{0} \sqrt{p_{{\rm p},k}} \phi_k[i] p\bigl(t-(i-1)T\bigr) \nonumber\\
    &\quad + \sum_{i=1}^{N_\text{d}} \sqrt{p_{{\rm d},k}} s_k[i] p\bigl(t-(i-1)T-\tau_kT\bigr),
\end{align}
where $p(\cdot)$ is the unit-energy pulse shape while $p_{{\rm p},k}$ and $p_{{\rm d},k}$ denote the pilot and data transmit powers, respectively.

The $\ell$th AP receives
\begin{align}\label{eq:rx_sig}
    \boldsymbol{y}_\ell(t) & = \sum \limits_{k = 1}^K \boldsymbol{h}_{\ell, k} x_k(t) + \boldsymbol{n}_\ell(t). 
\end{align}

\section{Uplink Reception: Synchronous Pilots and Asynchronous Data}
\label{sec:uplink_reception}


\subsection{Pilot Observations}

Upon pilot transmissions, each AP samples its MF output at the common times $t=(j-1)T$, with $j = -(N_\text{p}-1), \dots, 0$.
At the $\ell$th AP, that yields
\begin{align}\label{eq:pilot_mf_samples_sync}
    \boldsymbol{\mathsf{y}}_\ell[j]
    &=
    \boldsymbol{y}_\ell(t) * p^*(-t)\rvert_{t=(j-1)T}
\end{align}
Plugging \eqref{eq:tx_sig_syncpilot_asyncdata} and \eqref{eq:rx_sig} into \eqref{eq:pilot_mf_samples_sync}, 
\begin{align}\label{eq:pilot_obs_general_sync}
    \boldsymbol{\mathsf{y}}_\ell[j]
    &= \!
    \sum_{m=1}^{K}
    \sum_{i=-N_\text{p}+1}^{0} \!\!
    \sqrt{p_{{\rm p},m}}  \phi_m[i]
    q\bigl((j-i)T\bigr)\boldsymbol{h}_{\ell, m}\\
    &\quad \! + \!
    \sum_{m=1}^{K}
    \sum_{i=1}^{N_\text{d}}
    \sqrt{p_{{\rm d},m}}  s_m[i]
    q\bigl((j-i)T-\tau_mT\bigr)\boldsymbol{h}_{\ell, m}
    \nonumber+
    \boldsymbol{\mathsf{n}}_\ell[j], \notag
\end{align}
where
\begin{align}\label{eq:pulse_corr}
    q(t) = p(t) * p^*(-t).
\end{align}
The first term in \eqref{eq:pilot_obs_general_sync} is the synchronous pilot contribution, the second term is the interference caused by the asynchronous data leaking into the pilot observation window, and
\begin{align}
    \boldsymbol{\mathsf{n}}_\ell[j]
    &=
    \boldsymbol{n}_\ell(t) * p^*(-t)\rvert_{t=(j-1)T}
\end{align}
is the complex Gaussian  noise, satisfying
\begin{equation}\label{eq:noise_cov}
\mathbb{E} \! \left[ \boldsymbol{n}_\ell(t) \boldsymbol{n}_\ell^*(\tau)\right] = \sigma^2 \delta(t-\tau) \boldsymbol{I}_{N_\text{R}},
\end{equation}
where $\sigma^2$ is the noise power per AP antenna.
If $p(t)$ is a Nyquist or a rectangular pulse, then $q((j-i)T)=\delta_{i,j}$.
It follows that
\begin{align}\label{eq:pilot_obs_general_sync2}
    \boldsymbol{\mathsf{y}}_\ell[j]
    &= \!
    \sum_{m=1}^{K}
    \sqrt{p_{{\rm p},m}}  \phi_m[j]
   \boldsymbol{h}_{\ell, m}
    \\
    &\quad \! + \!
    \sum_{m=1}^{K}
    \sum_{i=1}^{N_\text{d}}
    \sqrt{p_{{\rm d},m}}  s_m[i]
    q\bigl((j-i)T-\tau_mT\bigr)\boldsymbol{h}_{\ell, m}
    \nonumber
    + 
    \boldsymbol{\mathsf{n}}_\ell[j].   \notag
\end{align}
Then, letting
\begin{align}\label{eq:pilot_stack}
    \boldsymbol{\mathsf{Y}}_\ell
    =
    \bigl[
    \boldsymbol{\mathsf{y}}_\ell[-N_\text{p}+1],
    \dots,
    \boldsymbol{\mathsf{y}}_\ell[0]
    \bigr]
    \in \mathbb{C}^{N_\text{R}\times N_\text{p}},
\end{align}
\begin{align}\label{eq:pilot_noise_stack}
    \boldsymbol{\mathsf{N}}_\ell
    =
    \bigl[
    \boldsymbol{\mathsf{n}}_\ell[-N_\text{p}+1],
    \dots,
    \boldsymbol{\mathsf{n}}_\ell[0]
    \bigr]
    \in \mathbb{C}^{N_\text{R}\times N_\text{p}},
\end{align}
and defining $\boldsymbol{\mathsf{A}}_{m}\in\mathbb{C}^{N_\text{p}\times N_\text{d}}$ with entries,
\begin{align}\label{eq:Ap_entries}
    [\boldsymbol{\mathsf{A}}_{m}]_{j,i}
    =
    q\bigl((j-i)T-\tau_mT\bigr),
\end{align}
where $j=-N_\text{p}+1,\dots,0$ and $i=1,\dots,N_\text{d}$, one can define the pilot observation matrix
\begin{align}\label{eq:pilot_matrix_model_sync_full}
    \boldsymbol{\mathsf{Y}}_\ell
    =
    \sum_{m=1}^{K} \!
    \sqrt{p_{{\rm p},m}} 
    \boldsymbol{h}_{\ell, m}\boldsymbol{\phi}_m^\top
    + \!
    \sum_{m=1}^{K}
    \sqrt{p_{{\rm d},m}} 
    \boldsymbol{h}_{\ell, m}
    (\boldsymbol{\mathsf{A}}_{m}\boldsymbol{s}_m)^\top
    +
    \boldsymbol{\mathsf{N}}_\ell .
\end{align}
Recalling \eqref{eq:orthogonal_pilots}, a sufficient statistic to estimate $\boldsymbol{h}_{\ell,k}$ is
\begin{align}
    \boldsymbol{\mathsf{z}}_{\ell, k}
   & =
    \frac{1}{N_\text{p}}
    \boldsymbol{\mathsf{Y}}_\ell \bar{\boldsymbol{\phi}}_k \\
 &   =
    \sqrt{p_{{\rm p},k}}\boldsymbol{h}_{\ell, k}
    +
    \boldsymbol{\eta}_{\ell, k},
    \label{eq:zpk_MMSE_model_final}
\end{align}
where
\begin{align}\label{eq:eta_lk}
    \boldsymbol{\eta}_{\ell, k}
    =
    \frac{1}{N_\text{p}}
    \sum_{m=1}^{K}
    \sqrt{p_{{\rm d},m}}
    \boldsymbol{h}_{\ell, m}
    (\boldsymbol{\mathsf{A}}_{m}\boldsymbol{s}_m)^\top
    \bar{\boldsymbol{\phi}}_k
    +
    \boldsymbol{\mathsf{w}}_{\ell, k}
\end{align}
subsumes both data leakage into the pilot observation interval and filtered noise, and $\bar{\boldsymbol{\phi}}$ denotes the conjugate of $\boldsymbol{\phi}$.


\subsection{Channel Estimation}

The true channel $\boldsymbol{h}_{\ell, k}$ can be expressed as 
$
    \boldsymbol{h}_{\ell, k} = \hat{\boldsymbol{h}}_{\ell, k} + \tilde{\boldsymbol{h}}_{\ell, k},
$
with $\hat{\boldsymbol{h}}_{\ell, k}$ being its estimate and $\tilde{\boldsymbol{h}}_{\ell, k}$ the estimation error \cite{4712720}.
The MMSE estimate obtained from $\boldsymbol{\mathsf{z}}_{\ell, k}$ is \cite[Sec. 2.7]{foundations2018}
\begin{align}
    \hat{\boldsymbol{h}}_{\ell, k}^{\rm MMSE}
    & =
    \mathbb{E} \! \left[
    \boldsymbol{h}_{\ell, k} \boldsymbol{\mathsf{z}}_{\ell, k}^*
    \right]
    \left(
    \mathbb{E} \! \left[
    \boldsymbol{\mathsf{z}}_{\ell, k} \boldsymbol{\mathsf{z}}_{\ell, k}^*
    \right]
    \right)^{-1}
    \boldsymbol{\mathsf{z}}_{\ell, k} \\
& =     \sqrt{p_{{\rm p},k}}
    \boldsymbol{R}_{\ell, k}
    \left(
    p_{{\rm p},k}\boldsymbol{R}_{\ell, k}
    +
    \boldsymbol{C}_{\boldsymbol{\eta}_{\ell, k}}
    \right)^{-1}
    \boldsymbol{\mathsf{z}}_{\ell, k},
    \label{eq:MMSE_estimate_syncpilot_final}  
\end{align}
where
\begin{align}\label{eq:disturbance_cov}
    \boldsymbol{C}_{\boldsymbol{\eta}_{\ell, k}}= \mathbb{E} \! \left[ \boldsymbol{\eta}_{\ell, k}\boldsymbol{\eta}_{\ell, k}^* 
    \right],
\end{align}
    which, because the data symbols are mutually independent and independent of the channels, is equal to (see Appendix~\ref{app:proof_disturbance_cov})
\begin{align}
    \boldsymbol{C}_{\boldsymbol{\eta}_{\ell, k}}
    &=
    \frac{1}{N_\text{p}^2}
    \sum_{m=1}^{K}
    p_{{\rm d},m}
    \boldsymbol{\phi}_k^\top
    \bar{\boldsymbol{\mathsf{A}}}_{m}
    \boldsymbol{\mathsf{A}}_{m}^\top
    \bar{\boldsymbol{\phi}}_k
    \boldsymbol{R}_{\ell, m}  +\frac{\sigma^2}{N_\text{p}}\boldsymbol{I}_{N_\text{R}}.
\end{align}

\subsection{Asynchronous Data Observations}

Each AP samples at the UE delays, $(j-1)T + \tau_kT$ for every $k$. Thus, the MF output at the $\ell$th AP associated with the $k$th UE is 
\begin{align}\label{eq:rx_mf}
    \boldsymbol{y}_{\ell, k}[j] &= \boldsymbol{y}_\ell(t) * p^{*}(-t)\rvert_{t=(j-1)T+\tau_k T} 
\end{align}
These samples provide sufficient statistics for the transmitted symbols \cite{verdu1989asynchronous}. Recalling $q(t)$ in \eqref{eq:pulse_corr}, let us define
\begin{align}
q_{k,m}(j-i) = q\big((j-i)T + (\tau_k - \tau_m)T\big) .
\end{align}
Plugging \eqref{eq:tx_sig_syncpilot_asyncdata} and \eqref{eq:rx_sig} into \eqref{eq:rx_mf},
\begin{align}
    \boldsymbol{y}_{\ell, k}[j] 
    &=
    \sum_{m=1}^{K}
    \sum_{i=1}^{N_\text{d}}
    \sqrt{p_{{\rm d},m}}  s_m[i]
    q\bigl((j-i)T+(\tau_k-\tau_m)T\bigr)\boldsymbol{h}_{\ell, m}
    \nonumber\\
    &\quad+
    \sum_{m=1}^{K}
    \sum_{i=-N_\text{p}+1}^{0}
   \!\! \sqrt{p_{{\rm p},m}}  \phi_m[i]
    q\bigl((j-i)T+\tau_kT\bigr)\boldsymbol{h}_{\ell, m}
    \nonumber\\
    &\quad+
    \boldsymbol{n}_{\ell, k}[j], 
    \label{eq:data_obs_before_sic_syncpilot_full}
\end{align}
where 
\begin{align}
    \boldsymbol{n}_{\ell, k}[j] 
    &= \int_{-\infty}^{\infty} \boldsymbol{n}_\ell(t) \, p^*\big(t - (j-1)T - \tau_k T\big)  \mathrm{d}t.
\end{align}
The first term in \eqref{eq:data_obs_before_sic_syncpilot_full} is the asynchronous data while the second term is residual pilot interference leaking into the data interval. To render the notation more compact, let us define
\begin{align}
    \boldsymbol{y}_{\ell, k}
    =
    \left[
    \boldsymbol{y}_{\ell, k}^{\top}[1],
    \boldsymbol{y}_{\ell, k}^{\top}[2],
    \dots,
    \boldsymbol{y}_{\ell, k}^{\top}[N_\text{d}]
    \right]^\top
    \in \mathbb{C}^{N_\text{R}N_\text{d}\times 1}
\end{align}
and similarly define $\boldsymbol{n}_{\ell, k}$. Further, let us define $\boldsymbol{Q}_{k,m}\in\mathbb{C}^{N_\text{d}\times N_\text{d}}$ with entries
\begin{align}\label{eq:Qd_entries_syncpilot_full}
    [\boldsymbol{Q}_{k,m}]_{j,i}
    =
    q\bigl((j-i)T+(\tau_k-\tau_m)T\bigr)
\end{align}
as well as $\boldsymbol{\mathsf{B}}_{k}\in\mathbb{C}^{N_\text{d}\times N_\text{p}}$ with entries
\begin{align}\label{eq:Pk_entries_syncpilot_full}
    [\boldsymbol{\mathsf{B}}_{k}]_{j,i}
    =
    q\bigl((j-i)T+\tau_kT\bigr).
\end{align}
Then,  \eqref{eq:data_obs_before_sic_syncpilot_full} can be compactly rewritten as
\begin{align}\label{eq:data_obs_matrix_before_sic_syncpilot_full}
    \boldsymbol{y}_{\ell, k}
    &=
    \sum_{m=1}^{K}
    \sqrt{p_{{\rm d},m}} 
    (\boldsymbol{I}_{N_\text{d}}\otimes \boldsymbol{h}_{\ell, m})
    \boldsymbol{Q}_{k,m}\boldsymbol{s}_m
    \nonumber\\
    &\quad+
    \sum_{m=1}^{K}
    \sqrt{p_{{\rm p},m}} 
    (\boldsymbol{I}_{N_\text{d}}\otimes \boldsymbol{h}_{\ell, m})
    \boldsymbol{\mathsf{B}}_{k}\boldsymbol{\phi}_m
    +
    \boldsymbol{n}_{\ell, k}.
\end{align}
The AP can estimate the pilot contribution as 
\begin{align}\label{eq:pilot_reconstruction_syncpilot_full}
    \widehat{\boldsymbol{\mathsf{r}}}_{\ell, k}
    =
    \sum_{m=1}^{K}
    \sqrt{p_{{\rm p},m}} 
    (\boldsymbol{I}_{N_\text{d}}\otimes \hat{\boldsymbol{h}}_{\ell, m})
   \boldsymbol{\mathsf{B}}_{k}\boldsymbol{\phi}_m
\end{align}
and subtract it from the observations, yielding
\begin{align}
    \bar{\boldsymbol{y}}_{\ell, k}
    =
    \boldsymbol{y}_{\ell, k}
    -
    \widehat{\boldsymbol{\mathsf{r}}}_{\ell, k}.
\end{align}
Using $\boldsymbol{h}_{\ell, m}= \hat{\boldsymbol{h}}_{\ell, m} +\tilde{\boldsymbol{h}}_{\ell, m},$ 
\begin{align}
\label{eq:post_cancel_model_syncpilot_full}
    \bar{\boldsymbol{y}}_{\ell, k}
    &=
    \sum_{m=1}^{K}
    \sqrt{p_{{\rm d},m}} 
    (\boldsymbol{I}_{N_\text{d}}\otimes \hat{\boldsymbol{h}}_{\ell, m})
    \boldsymbol{Q}_{k,m}\boldsymbol{s}_m
    \nonumber\\
    &\quad+
    \sum_{m=1}^{K}
    \sqrt{p_{{\rm d},m}} 
    (\boldsymbol{I}_{N_\text{d}}\otimes \tilde{\boldsymbol{h}}_{\ell, m})
    \boldsymbol{Q}_{k,m}\boldsymbol{s}_m
    \nonumber\\
    &\quad+
    \sum_{m=1}^{K}
    \sqrt{p_{{\rm p},m}} 
    (\boldsymbol{I}_{N_\text{d}}\otimes \tilde{\boldsymbol{h}}_{\ell, m})
    \boldsymbol{\mathsf{B}}_{k}\boldsymbol{\phi}_m
    +
    \boldsymbol{n}_{\ell, k}.
\end{align}
Defining the stacked vector
\begin{align}
    \bar{\boldsymbol{y}}_{\ell}
    =
    \left[
    \bar{\boldsymbol{y}}_{\ell, 1}^{\top},
    \bar{\boldsymbol{y}}_{\ell, 2}^{\top},
    \dots,
    \bar{\boldsymbol{y}}_{\ell, K}^{\top}
    \right]^\top
    \in \mathbb{C}^{KN_\text{R}N_\text{d}\times 1}
\end{align}
along with 
\begin{align}
    \boldsymbol{s} &
    =
    \left[
    \boldsymbol{s}_1^\top,\dots,\boldsymbol{s}_K^\top
    \right]^\top,
    \\
    \boldsymbol{n}_{\ell}
    & =
    \left[
    \boldsymbol{n}_{\ell,1}^{\top},
    \dots,
    \boldsymbol{n}_{\ell, K}^{\top}
    \right]^\top,
\end{align}
the stacked vector $\bar{\boldsymbol{y}}_{\ell}$ can be compactly expressed as
\begin{align}
\label{eq:stacked_post_cancel_model_syncpilot_full}
    \bar{\boldsymbol{y}}_{\ell}
    =
    \boldsymbol{G}_{\ell}\boldsymbol{s}
    + \underbrace{
    \boldsymbol{r}^{{\rm res}}_{\ell}
    +
    \boldsymbol{\mathsf{r}}^{{\rm res}}_{\ell}
    +
    \boldsymbol{n}_{\ell}}_{\boldsymbol{\eta}_\ell},
\end{align}
where $\boldsymbol{G}_{\ell}\in\mathbb{C}^{KN_\text{R}N_\text{d}\times KN_\text{d}}$ is the asynchronous data matrix
\begin{align}
    \boldsymbol{G}_{\ell}
    =
    \begin{bmatrix}
        \boldsymbol{G}_{\ell, 1,1} & \boldsymbol{G}_{\ell, 1,2} & \cdots & \boldsymbol{G}_{\ell 1,K} \\
        \boldsymbol{G}_{\ell, 2,1} & \boldsymbol{G}_{\ell, 2,2} & \cdots & \boldsymbol{G}_{\ell, 2,K} \\
        \vdots & \vdots & \ddots & \vdots \\
        \boldsymbol{G}_{\ell, K,1} & \boldsymbol{G}_{\ell, K,2} & \cdots & \boldsymbol{G}_{\ell, K,K}
    \end{bmatrix},
\end{align}
with
\begin{align}
\label{eq:Gd_block_def_syncpilot_full}
    \boldsymbol{G}_{\ell, k,m}
    =
    \sqrt{p_{{\rm d},m}} 
    (\boldsymbol{I}_{N_\text{d}}\otimes \hat{\boldsymbol{h}}_{\ell, m})
    \boldsymbol{Q}_{k,m}.
\end{align}
The residual error on the data portion of the channel induced by imperfect channel estimation is
\begin{align}
\label{eq:data_residual_block_syncpilot_full}
    \boldsymbol{r}^{{\rm res}}_{\ell, k}
    =
    \sum_{m=1}^{K}
    \sqrt{p_{{\rm d},m}} 
    (\boldsymbol{I}_{N_\text{d}}\otimes \tilde{\boldsymbol{h}}_{\ell, m})
    \boldsymbol{Q}_{k,m}\boldsymbol{s}_m,
\end{align}
and the residual pilot interference is
\begin{align}\label{eq:pilot_residual_block_syncpilot_full}
    \boldsymbol{\mathsf{r}}^{{\rm res}}_{\ell, k}
    =
    \sum_{m=1}^{K}
    \sqrt{p_{{\rm p},m}} 
    (\boldsymbol{I}_{N_\text{d}}\otimes \tilde{\boldsymbol{h}}_{\ell, m})
    \boldsymbol{\mathsf{B}}_{k}\boldsymbol{\phi}_m.
\end{align}

The channel estimation errors of different UEs can be regarded as uncorrelated---the spurious effects induced by asynchronous data leakages into the pilot interval are rather negligible---
and the cross-covariance between
$\boldsymbol{\mathsf{r}}^{{\rm res}}_{\ell, k}$ and
$\boldsymbol{\mathsf{r}}^{{\rm res}}_{\ell, k'}$ at the $\ell$th AP is then
(see Appendix~\ref{app:proof_residual_cov})
\begin{align}\label{eq:pilot_residual_cross_covariance_syncpilot_full}
    \boldsymbol{\mathsf{C}}^{{\rm res}}_{\ell,k,k'}
    &=
    \mathbb{E} \! \left[
    \boldsymbol{\mathsf{r}}^{{\rm res}}_{\ell, k}
    \bigl(\boldsymbol{\mathsf{r}}^{{\rm res}}_{\ell, k'}\bigr)^*
    \right]
    \\
    &=
    \sum_{m=1}^{K}
    p_{{\rm p},m}
    \left(
    \boldsymbol{\mathsf{B}}_{k}\boldsymbol{\phi}_m
    \boldsymbol{\phi}_m^{*}
    \boldsymbol{\mathsf{B}}_{k'}^*
    \right)
    \otimes
    \boldsymbol{C}_{\tilde{\boldsymbol{h}}_{\ell, m}},
\end{align}
where
$
    \boldsymbol{C}_{\tilde{\boldsymbol{h}}_{\ell, m}}
    =
    \mathbb{E} \big[
    \tilde{\boldsymbol{h}}_{\ell, m}
    \tilde{\boldsymbol{h}}_{\ell, m}^*
    \big] .
$
%
A special case of the above is the autocovariance matrix 
\begin{align}
    \boldsymbol{\mathsf{C}}_{\boldsymbol{\mathsf{r}}_\ell}^{\rm res}
    =
    \begin{bmatrix}
        \boldsymbol{\mathsf{C}}^{{\rm res}}_{\ell,1,1}
        &
        \boldsymbol{\mathsf{C}}^{{\rm res}}_{\ell,1,2}
        &
        \cdots
        &
        \boldsymbol{\mathsf{C}}^{{\rm res}}_{\ell,1,K}
        \\
        \boldsymbol{\mathsf{C}}^{{\rm res}}_{\ell,2,1}
        &
        \boldsymbol{\mathsf{C}}^{{\rm res}}_{\ell,2,2}
        &
        \cdots
        &
        \boldsymbol{\mathsf{C}}^{{\rm res}}_{\ell,2,K}
        \\
        \vdots & \vdots & \ddots & \vdots
        \\
       \boldsymbol{\mathsf{C}}^{{\rm res}}_{\ell,K,1}
        &
       \boldsymbol{\mathsf{C}}^{{\rm res}}_{\ell,K,2}
        &
        \cdots
        &
       \boldsymbol{\mathsf{C}}^{{\rm res}}_{\ell,K,K}
    \end{bmatrix}.
\end{align}

Analogously,
the cross-covariance between
$\boldsymbol{r}^{{\rm res}}_{\ell, k}$ and
$\boldsymbol{r}^{{\rm res}}_{\ell, k'}$ at the $\ell$th AP can be seen to equal
\begin{align}
\label{eq:data_residual_cross_covariance_syncpilot_full}
    \boldsymbol{C}^{{\rm res}}_{\ell,k,k'}
    &=
    \sum_{m=1}^{K}
    p_{{\rm d},m}
    \left(
    \boldsymbol{Q}_{k,m}
    \boldsymbol{Q}_{k',m}^*
    \right)
    \otimes
    \boldsymbol{C}_{\tilde{\boldsymbol{h}}_{\ell,m}}.
\end{align}

The covariance of $\boldsymbol{r}^{{\rm res}}_{\ell}$, denoted by $\boldsymbol{C}_{\boldsymbol{r}_\ell}^{\rm res}$, is the $K\times K$ block matrix whose $(k,k')$th block is $\boldsymbol{C}^{{\rm res}}_{\ell,k,k'}$, analogously to $\boldsymbol{\mathsf{C}}_{\boldsymbol{\mathsf{r}}_\ell}^{\rm res}$.

Importantly, the entries of the noise vector are not independent because of the asynchronicity. 
The covariance between  $\boldsymbol{n}_{\ell, k}[j]$ and $\boldsymbol{n}_{\ell, m}[i]$ is given by~\eqref{eq:noise_integral}, atop the next page.
\begin{figure*}[!t]
\centering
\begin{minipage}{0.98\textwidth}
\begin{align} \label{eq:noise_integral}
\mathbb{E} \Bigg[
\boldsymbol{n}_{\ell, k}[j]
\boldsymbol{n}^*_{\ell, m}[i]
\Bigg]
&=
\mathbb{E}\Bigg[
\biggl(
\int_{-\infty}^{\infty}
\boldsymbol{n}_\ell(t)
p^*(t - (j-1)T - \tau_k T)
\mathrm{d}t
\biggr)
\biggl(
\int_{-\infty}^{\infty}
\boldsymbol{n}_\ell(\tau)
p^*(\tau - (i-1)T - \tau_m T)
\mathrm{d}\tau
\biggr)^{*}
\Bigg] \nonumber\\
&=
\int_{-\infty}^{\infty} \int_{-\infty}^{\infty}
\mathbb{E} \! \left[
\boldsymbol{n}_\ell(t)\boldsymbol{n}^*_\ell(\tau)
\right] 
p^* (t - (j-1)T - \tau_k T )
p\left(\tau - (i-1)T - \tau_m T\right)
\mathrm{d}t  \mathrm{d}\tau 
\end{align}

\vspace{-0.5em}
\noindent\rule{\textwidth}{0.4pt}
\vspace{-0.8em}

\end{minipage}
\vspace{-1.5em}
\end{figure*}
Using \eqref{eq:noise_cov} and simplifying,
\begin{align}
\mathbb{E} \bigg[
\boldsymbol{n}_{\ell, k}[j]
\boldsymbol{n}^*_{\ell, m}[i]\bigg]
&= \sigma^2 q_{k,m}(j-i)\boldsymbol{I}_{N_{\text{R}}} \\
& = \sigma^2 [Q_{k,m}]_{j,i}\boldsymbol{I}_{N_{\text{R}}},
\end{align}
whereby the covariance between $\boldsymbol{n}_{\ell, k}$ and $\boldsymbol{n}_{\ell, m}$ is
\begin{equation}
\mathbb{E} \! \left[\boldsymbol{n}_{\ell, k} \boldsymbol{n}_{\ell, m}^*\right] = \sigma^2 \boldsymbol{Q}_{k,m} \otimes \boldsymbol{I}_{N_{\text{R}}},
\end{equation}
where $\boldsymbol{Q}_{k,m}$ captures the pulse overlaps.

By stacking all users and time samples into the full noise vector $\boldsymbol{n}_\ell \in \mathbb{C}^{KN_\text{d}N_{\text{R}} \times 1}$, the overall noise covariance becomes 
\begin{align}
\mathbb{E} \! \left[\boldsymbol{n}_\ell \boldsymbol{n}_\ell^*\right]
&= \sigma^2
\begin{bmatrix}
\boldsymbol{Q}_{1,1} \otimes \boldsymbol{I}_{N_{\text{R}}} & \cdots & \boldsymbol{Q}_{1,K} \otimes \boldsymbol{I}_{N_{\text{R}}} \\
\vdots & \ddots & \vdots \\
\boldsymbol{Q}_{K,1} \otimes \boldsymbol{I}_{N_{\text{R}}} & \cdots & \boldsymbol{Q}_{K,K} \otimes \boldsymbol{I}_{N_{\text{R}}} \notag
\end{bmatrix} \\
&= \sigma^2 \left(
\begin{bmatrix}
\boldsymbol{Q}_{1,1} & \boldsymbol{Q}_{1,2} & \cdots & \boldsymbol{Q}_{1,K} \\
\boldsymbol{Q}_{2,1} & \boldsymbol{Q}_{2,2} & \cdots & \boldsymbol{Q}_{2,K} \\
\vdots & \vdots & \ddots & \vdots \\
\boldsymbol{Q}_{K,1} & \boldsymbol{Q}_{K,2} & \cdots & \boldsymbol{Q}_{K,K}
\end{bmatrix}
\otimes \boldsymbol{I}_{N_{\text{R}}}
\right) \nonumber\\
&= \sigma^2 \boldsymbol{Q}\otimes \boldsymbol{I}_{N_{\text{R}}}.
\end{align}
Stacking the observations for the $L$ APs, with the pilot contributions (reconstructed using the estimated channels) subtracted,
\begin{align}
    \bar{\boldsymbol{y}}
    =
    \left[
    \bar{\boldsymbol{y}}^\top_{1},
    \bar{\boldsymbol{y}}^\top_{2},
    \dots,
    \bar{\boldsymbol{y}}^\top_{L}
    \right]^\top
    \in \mathbb{C}^{LKN_\text{d}N_{\text{R}} \times 1},
\end{align}
with $\bar{\boldsymbol{y}}_{\ell}$ as per \eqref{eq:stacked_post_cancel_model_syncpilot_full}. Then,
\begin{align}
\label{eq:matrix_model_syncpilot_final}
    \bar{\boldsymbol{y}}
    &=
    \begin{pmatrix}
        \boldsymbol{G}_{1} \\
        \boldsymbol{G}_{2} \\
        \vdots \\
        \boldsymbol{G}_{L}
    \end{pmatrix}
    \boldsymbol{s}
    +
    \begin{pmatrix}
        \boldsymbol{r}^{{\rm res}}_{1} \\
        \boldsymbol{r}^{{\rm res}}_{2} \\
        \vdots \\
        \boldsymbol{r}^{{\rm res}}_{L}
    \end{pmatrix}
    +
    \begin{pmatrix}
        \boldsymbol{\mathsf{r}}^{{\rm res}}_{1} \\
         \boldsymbol{\mathsf{r}}^{{\rm res}}_{2} \\
        \vdots \\
         \boldsymbol{\mathsf{r}}^{{\rm res}}_{L}
    \end{pmatrix}
    +
    \begin{pmatrix}
        \boldsymbol{n}_{1} \\
        \boldsymbol{n}_{2} \\
        \vdots \\
        \boldsymbol{n}_{L}
    \end{pmatrix}\\
    &
    =
    \boldsymbol{G}\boldsymbol{s}
    +
    \underbrace{
    \boldsymbol{r}^{{\rm res}}
    +
     \boldsymbol{\mathsf{r}}^{{\rm res}}
    +
    \boldsymbol{n}}_{\boldsymbol{\eta}}, \label{BCN}
\end{align}
where $\boldsymbol{s} \in \mathbb{C}^{KN_\text{d}\times 1}$ is the stacked data vector whereas $\boldsymbol{G}\in\mathbb{C}^{LKN_\text{d}N_\text{R} \times KN_\text{d}}$ is the
channel-estimate-based asynchronous data matrix and $\boldsymbol{\eta}$ contains residual errors, residual pilot interference, and colored noise.
The covariance matrices of $\boldsymbol{r}^{{\rm res}}$ and $ \boldsymbol{\mathsf{r}}^{{\rm res}}$ are block diagonal 
because the channel-estimation errors are independent across APs, and
\begin{align}
\label{eq:central_noise_covariance_all_aps}
    \boldsymbol{C}_{\boldsymbol{n}}
    =
    \sigma^2
    \left(
    \boldsymbol{I}_{L}\otimes
    \boldsymbol{Q}\otimes
    \boldsymbol{I}_{N_\text{R}}
    \right).
\end{align}

\subsection{User- and AP-Centric Subsets}
\label{subsec:user_ap_subsets}

Let $d_{\ell, k}\in\{0,1\}$ indicate whether AP $\ell$ participates in the reception of UE $k$. The ensuing user-centric AP subset and AP-centric UE subsets are, respectively,
\begin{align}
    \mathcal{M}_k
    &=
    \left\{
    \ell\in\{1,\ldots,L\}:d_{\ell, k}=1
    \right\}                                                   \\
    \mathcal{A}_{\ell}
    &=
    \left\{
    k\in\{1,\ldots,K\}:d_{\ell, k}=1
    \right\}
    =
    \left\{
    k:\ell\in\mathcal{M}_k
    \right\}.
\end{align}
Thus, $\mathcal{M}_k$ contains the APs whose observations are used to detect UE $k$ while $\mathcal{A}_{\ell}$ contains the UEs processed by AP $\ell$. The association is constructed using the large-scale channel gains, i.e., each UE is assigned to its strongest APs while respecting
\begin{align}
    \left|\mathcal{M}_k\right|
    \leq
    M_{\max}
    \qquad
    \left|\mathcal{A}_{\ell}\right|
    \leq
    S_{\max}.
\end{align}
Signals transmitted by UEs outside $\mathcal{A}_{\ell}$ constitute interference.

\subsection{Central Reception}
\label{subsec:central_reception}

When AP observations are forwarded to a central processing unit,
symbol $s_k[j]$ is detected as
\begin{align}
    \hat{s}_k[j]
    =
    \left(\boldsymbol{v}_{k}^{\rm cen}[j]\right)^{\!*}
    \bar{\boldsymbol{y}}
\end{align}
with $\boldsymbol{\eta}$ in (\ref{BCN}) as the disturbance.

Let $i=(k-1)N_\text{d}+j$ be the position of $s_k[j]$ in the
stacked data vector, while $\boldsymbol{g}_i$ is the $i$th column of
$\boldsymbol{G}$ and $\boldsymbol{G}_{-i}$ denotes
$\boldsymbol{G}$ with this column removed. The centralized MMSE
combiner is
\begin{align}
\label{eq:central_combiner_compact}
    \boldsymbol{v}_{k}^{\rm cen}[j]
    &=
    \left(
    \boldsymbol{G}_{-i}\boldsymbol{G}_{-i}^*
    +
    \boldsymbol{C}_{\boldsymbol{\eta}}
    \right)^{-1}
    \boldsymbol{g}_i
\end{align}    
and its SINR is
\begin{align}
\label{eq:central_processing_sinr}
    \mathrm{SINR}_{k}^{\rm cen}[j]
    &=
    \frac{
    \left|
    \big(\boldsymbol{v}_{k}^{\rm cen}[j]\big)^*
    \boldsymbol{g}_i
    \right|^2
    }{
    \left\|
    \big(\boldsymbol{v}_{k}^{\rm cen}[j]\big)^*
    \boldsymbol{G}_{-i}
    \right\|^2
    +
    \big(\boldsymbol{v}_{k}^{\rm cen}[j]\big)^*
    \boldsymbol{C}_{\boldsymbol{\eta}}
    \boldsymbol{v}_{k}^{\rm cen}[j]
    }.
\end{align}
Although the above combiner is time-varying, its variability is only significant in the vicinity of the frame boundaries, and its value for inner symbols can be applied over the entire frame with little penalty (see the performance evaluation in Sec. \ref{sec:simulation_results}).
We denote by $\boldsymbol{v}_{k}^{\rm cen}$, with the symbol index dropped, the time-invariant combiner obtained by evaluating \eqref{eq:central_combiner_compact} for the middle symbol $j_0=\lceil N_\text{d}/2\rceil$ and applying it to every symbol in the frame. To further reduce computational complexity, we introduce a windowed combiner. Because the pulse overlap decays with the separation between symbols, samples far from symbol $j$ carry little information about $s_k[j]$. The combiner dimension can therefore be reduced by retaining, on each sampling branch, only the MF samples with indices $j'\in\{j-W,\ldots,j+W\}$, where $W$ is the window half-width. The combiner dimension then scales with $2W+1$ rather than with $N_\text{d}$.

To incorporate user-centric processing, let $\boldsymbol{S}_k$ select the observation blocks of the APs in $\mathcal{M}_k$.
The user-centric combiner and SINR follow directly from
\eqref{eq:central_combiner_compact}-\eqref{eq:central_processing_sinr}
after replacing
$\boldsymbol{g}_i$, $\boldsymbol{G}_{-i}$, and
$\boldsymbol{C}_{\boldsymbol{\eta}}$ by
$\boldsymbol{S}_k\boldsymbol{g}_i$,
$\boldsymbol{S}_k\boldsymbol{G}_{-i}$, and
\begin{align}
    \boldsymbol{C}_{\boldsymbol{\eta},k}
    =
    \boldsymbol{S}_k
    \boldsymbol{C}_{\boldsymbol{\eta}}
    \boldsymbol{S}_k^*.
\end{align}
With that, the
MMSE receiver is applied only to the AP observations relevant to UE
$k$.

\subsection{Local Processing}
\label{subsec:local_processing}

In local processing, AP $\ell$ uses its own observation
\begin{align}
    \bar{\boldsymbol{y}}_{\ell}
    =
    \boldsymbol{G}_{\ell}\boldsymbol{s}
    +
    \boldsymbol{\eta}_{\ell}.
\end{align}
With $i$ again the position of $s_k[j]$ within $\boldsymbol{s}$, let $\boldsymbol{g}_{\ell,i}$ be the column $i$ of $\boldsymbol{G}_{\ell}$ and let
$\boldsymbol{G}_{\ell,-i}$ be obtained by removing that
column. If $\ell\in\mathcal{M}_k$, then AP $\ell$ employs
\begin{align}
\label{eq:local_combiner}
    \boldsymbol{v}_{\ell, k}[j]
    =
    \left(
    \boldsymbol{G}_{\ell,-i}
    \boldsymbol{G}_{\ell,-i}^*
    +
    \boldsymbol{C}_{\boldsymbol{\eta},\ell}
    \right)^{-1}
    \boldsymbol{g}_{\ell,i}
\end{align}
to form the scalar soft estimate
$z_{\ell, k}[j]=\boldsymbol{v}_{\ell, k}[j]^*
\bar{\boldsymbol{y}}_{\ell}$.
Then, the central processing unit combines the soft estimates as
\begin{align}
\label{eq:local_cpu_combination}
    \hat{s}_k[j]
    =
    \sum_{\ell\in\mathcal{M}_k}z_{\ell, k}[j].
\end{align}
The resulting SINR is
\begin{align}
\label{eq:local_processing_sinr}
    \mathrm{SINR}_{k}^{\rm loc}[j]
    =
    \frac{
    \left|
    \displaystyle\sum_{\ell\in\mathcal{M}_k}
    \boldsymbol{v}^*_{\ell, k}[j]
    \boldsymbol{g}_{\ell,i}
    \right|^2
    }{
    \left\|
    \displaystyle\sum_{\ell\in\mathcal{M}_k}
    \boldsymbol{v}^*_{\ell, k}[j]
    \boldsymbol{G}_{\ell,-i}
    \right\|^2
    \! + \!
    \displaystyle\sum_{\ell\in\mathcal{M}_k}
    \boldsymbol{v}^*_{\ell, k}[j]
    \boldsymbol{C}_{\boldsymbol{\eta},\ell}
    \boldsymbol{v}_{\ell, k}[j]
    }.
\end{align}
As in the centralized case, we use 
$\boldsymbol{v}_{\ell,k}$ with the symbol dependence dropped to indicate the combiner evaluated for the middle symbol in the frame; this is the time-invariant alternative to $\boldsymbol{v}_{\ell, k}[j]$.

\subsection{Performance Measure}

With the proviso of ergodicity, the spectral efficiency (in bits/s/Hz) achieved by the $k$th UE is
\begin{equation}
\mathrm{SE}_k
=
\frac{1}{N_\text{d}+N_\text{p}}
\sum_{j=1}^{N_\text{d}}
\mathbb{E}\big[
\log_2 (1 + \mathrm{SINR}_k[j] )
\big] .
\end{equation}
For fixed $N_\text{d}$ and $N_\text{p}$, the spectral efficiency is monotonic in the SINR. Hence, any optimization of the former can be more conveniently conducted on the latter. However, if the optimization does involve the pilot overhead, then it must take place directly on the spectral efficiency.

\subsection{Codebook-Limited and Reduced Sampling}
\label{subsec:codebook}

To prevent the number of sampling branches from growing with the
number of UEs, the intentional delays are restricted to a finite
codebook
\begin{equation}
    \mathcal{T}
    =
    \{\delta_1,\ldots,\delta_Q\},
\end{equation}
where $Q$ is independent of $K$. A uniform normalized codebook is
\begin{equation}
    \mathcal{T}
    =
    \left\{
    0,\frac{1}{Q},\ldots,\frac{Q-1}{Q}
    \right\}.
\end{equation}
The $\ell$th AP only samples the distinct delays assigned to its AP-centric
UE subset,
\begin{align}
    \mathcal{T}_{\ell}
    &=
    \{\tau_k:k\in\mathcal{A}_{\ell}\}
    \subseteq\mathcal{T}
\end{align}
with
\begin{align}
    |\mathcal{T}_{\ell}|
    & \leq
    \min\{|\mathcal{A}_{\ell}|,Q\} \\
    & \leq
    \min\{S_{\max},Q\}.
\end{align}
If
$\mathcal{T}_{\ell}
=\{\delta_{\ell,1},\ldots,\delta_{\ell, |\mathcal{T}_{\ell}|}\}$,
the corresponding matched-filter sampling instants are
$(j-1)T+\delta_{\ell,q}T$, for $q=1,\ldots, |\mathcal{T}_{\ell}|$.

After stacking the sampled delay branches and all data-symbol positions, let $\check{\boldsymbol{y}}_{\ell}\in
\mathbb{C}^{ |\mathcal{T}_{\ell}| N_{\mathrm d}N_{\mathrm R}}$ denote the
post-subtraction observation at AP $\ell$.
Let $\boldsymbol{s}_{\mathcal A_\ell}$ stack the $N_{\mathrm d}$ data
symbols of the UEs in $\mathcal A_\ell$, and let
$\check{\boldsymbol G}_{\ell,\mathcal A_\ell} \in\mathbb C^{ |\mathcal{T}_{\ell}| N_{\mathrm d}N_{\mathrm R}
\times |\mathcal A_\ell|N_{\mathrm d}}$ denote their effective
channel matrix, constructed using the channel estimates. The local
observation can then be written as
\begin{align}
\label{eq:reduced_local_model}
\check{\boldsymbol y}_\ell
=
\check{\boldsymbol G}_{\ell,\mathcal A_\ell}
\boldsymbol s_{\mathcal A_\ell}
+
\boldsymbol\eta_\ell^{\rm eff} .
\end{align}
The effective disturbance comprises filtered noise, channel estimation residuals, and interference from UEs outside $\mathcal A_\ell$.
The local combiner and SINR follow from \eqref{eq:local_combiner} and \eqref{eq:local_processing_sinr} after using the reduced channel matrix and its effective disturbance covariance. Since $ |\mathcal{T}_{\ell}|$ and $|\mathcal{A}_{\ell}|$ are bounded by constants independent of $K$, both the per-AP sampling requirement and the retained local receiver dimension remain bounded as the network grows.

\section{Interference-Based Surrogate Problem}
\label{sec:interference_surrogate}


The received signal model evidences that intentional delays shape the coupling among users across the MF observations. This can be exploited to improve the separation of
users that strongly interact at their serving APs.

\subsection{Multiuser Interference}

Recall that the asynchronous data observation in \eqref{eq:data_obs_before_sic_syncpilot_full} contains two terms. The first term is the asynchronous data component, where the coupling between the sampling branch of UE $k$ and the transmitted symbols of UE $m$ depends on $\tau_k-\tau_m$. The second term is the residual pilot interference that leaks into the data interval. As pilots are transmitted without user-specific intentional delays, this term depends on $\tau_k$, but not on $\tau_k-\tau_m$. And, as the pilot interference is caused by residual errors in channel estimation, it is weak. Thus, in the
delay assignment problem, only the asynchronous data term is considered to
construct the pairwise delay-dependent objective.

When detecting UE $k$, the $j$th data sample of the asynchronous data contribution of UE $m\neq k$ at AP $\ell$ is
\begin{equation}
\boldsymbol{z}_{\ell,k\leftarrow m}[j]
=
\sum_{i=1}^{N_{\rm d}}
\sqrt{p_{{\rm d},m}} s_m[i]
q\left((j-i)T+(\tau_k-\tau_m)T\right)
\boldsymbol{h}_{\ell, m}.
\notag 
\end{equation}
Using $\boldsymbol{h}_{\ell, m}\sim\mathcal{CN}(\boldsymbol{0},\boldsymbol{R}_{\ell, m})$ and $\mathbb{E}\{\|\boldsymbol{h}_{\ell, m}\|^2\}=\operatorname{tr}(\boldsymbol{R}_{\ell, m})$,
\begin{align}
\E\!\left[
\left\|
\boldsymbol{z}_{\ell,k\leftarrow m}[j]
\right\|^2
\right]
&=
p_{{\rm d},m}
\operatorname{tr}(\boldsymbol{R}_{\ell, m})
\label{eq:expected_pairwise_data_interference_exact}
\\
&\quad \cdot \sum_{i=1}^{N_{\rm d}}
\left|
q\left((j-i)T+(\tau_k-\tau_m)T\right)
\right|^2 .\notag
\end{align}
As the pulse-shape is truncated to an interval $[-U,\ldots,U]$,
\begin{equation}
\E \! \left[
\left\|
\boldsymbol{z}_{\ell,k\leftarrow m}[j]
\right\|^2
\right]
=
p_{{\rm d},m}\operatorname{tr}(\boldsymbol{R}_{\ell, m})
\Gamma(\tau_k-\tau_m),
\label{eq:pairwise_data_interference_upper_bound}
\end{equation}
where
\begin{equation}
\Gamma(\tau_k-\tau_m)
=
\sum_{u=-U}^{U}
\left|
q\left((u+\tau_k-\tau_m)T\right)
\right|^2.
\label{eq:pulse_overlap_function_data}
\end{equation}

Then, per~\eqref{eq:pairwise_data_interference_upper_bound}, the asynchronous data-interference energy from UE $m$ to
the sampling branch of UE $k$ at AP $\ell$ is equal to the product of the transmit data power, the large-scale channel gain, the number of AP antennas, and a relative-delay-dependent pulse-overlap factor.

\subsection{Aggregate Interference}

Adding \eqref{eq:pairwise_data_interference_upper_bound} over the APs that serve UE $k$ and, for the sake of compactness, letting $\boldsymbol{\tau} = [\tau_1, \dots, \tau_K]^\top$,
the asynchronous data interference from UE $m$ into UE $k$ is
\begin{equation}
I_{k\leftarrow m}(\boldsymbol{\tau})
=
p_{{\rm d},m} \!
\sum_{\ell\in\mathcal M_k} \!
\operatorname{tr}(\boldsymbol{R}_{\ell, m})
\Gamma(\tau_k-\tau_m),
\label{eq:directed_data_interference_bound}
\end{equation}
which, defining the directed large-scale data-interference coefficient
\begin{equation}
a_{k\leftarrow m}
=
p_{{\rm d},m}
\sum_{\ell\in\mathcal M_k}
\operatorname{tr}(\boldsymbol{R}_{\ell, m}),
\label{eq:directed_data_interference_weight}
\end{equation}
becomes
\begin{equation}
I_{k\leftarrow m}(\boldsymbol{\tau})
=
a_{k\leftarrow m}
\Gamma(\tau_k-\tau_m).
\label{eq:directed_data_bound_compact}
\end{equation}

The total delay-dependent asynchronous data interference
affecting UE $k$ is thus
\begin{equation}
I_{k}(\boldsymbol{\tau})
=
\sum_{\substack{m=1\\m\neq k}}^{K}
a_{k\leftarrow m}
\Gamma(\tau_k-\tau_m).
\label{eq:user_data_interference_bound}
\end{equation}
Summing over all UEs gives
\begin{align}
\sum_{k=1}^{K}I_{k}(\boldsymbol{\tau})
&=
\sum_{k=1}^{K}
\sum_{\substack{m=1\\m\neq k}}^{K}
a_{k\leftarrow m}
\Gamma(\tau_k-\tau_m).
\label{eq:sum_directed_data_bound}
\end{align}
For a symmetric pulse autocorrelation,
$
\Gamma(\Delta)=\Gamma(-\Delta),
$
the sum in \eqref{eq:sum_directed_data_bound} is equal to
\begin{equation}
\sum_{m<k}
c_{m,k}
\Gamma(\tau_m-\tau_k),
\label{eq:pairwise_data_bound_objective}
\end{equation}
where
\begin{align}
c_{m,k}
& =
a_{m\leftarrow k}
+
a_{k\leftarrow m} . 
\label{eq:pairwise_data_weight_general}
\end{align}
In lieu of the unwieldy maximization of the spectral efficiency, we solve the interference-based surrogate problem
\begin{equation}
\min_{\tau_1,\ldots,\tau_K\in\mathcal T}
\sum_{m<k}
c_{m,k}
\Gamma(\tau_m-\tau_k).
\label{eq:bound_minimization_data_delay_problem}
\end{equation}

\subsection{Alternative Justification for the Surrogate Problem}

Recall that $\mathcal{M}_k=\{\ell_{k,1},\ldots,\ell_{k,M_k}\}$. The observation
on sampling branch $k$, stacked over the APs serving UE $k$, is
\begin{equation}
\boldsymbol{y}_{\mathcal{M}_k, k}[j]
=
\left[
\boldsymbol{y}_{\ell_{k,1},k}^{\top}[j],
\ldots,
\boldsymbol{y}_{\ell_{k,M_k},k}^{\top}[j]
\right]^{\top}.
\label{eq:stacked_observation_surrogate_new}
\end{equation}
The corresponding desired-channel vector, stacked over the serving APs,
is
\begin{equation}
\boldsymbol{h}_{\mathcal{M}_k, k}
=
\left[
\boldsymbol{h}_{\ell_{k,1},k}^{\top},
\ldots,
\boldsymbol{h}_{\ell_{k,M_k},k}^{\top}
\right]^{\top}
\in\mathbb{C}^{M_kN_{\mathrm R}\times 1}.
\label{eq:stacked_desired_channel_surrogate_new}
\end{equation}

As $q(0)=1$ and $q(nT)=0$ for nonzero integer $n$, the desired contribution of symbol $s_k[j]$ on the sampling branch of UE $k$ is
$
\sqrt{p_{{\rm d},k}}s_k[j]\boldsymbol{h}_{\mathcal{M}_k, k}
$.
Hence,
\begin{equation}
\boldsymbol y_{\mathcal M_k,k}[j]
=
\sqrt{p_{{\rm d},k}}s_k[j]\boldsymbol h_{\mathcal M_k,k}
+ \!
\sum_{\substack{m=1\\m\neq k}}^K \!
\boldsymbol z_{k\leftarrow m}[j]
+
\boldsymbol u_k[j],
\label{eq:stacked_decomposition_surrogate_new}
\end{equation}
where $\boldsymbol u_k[j]$ collects residual pilot leakage and noise.
Denoting by $\boldsymbol{H}$ the collection of all channel vectors $\{\boldsymbol{h}_{\ell,k}\}$, for an arbitrary linear combiner $\boldsymbol v_k$,
the conditional output SINR is
\begin{equation}
\gamma_k(\boldsymbol\tau,\boldsymbol v_k)
=
\frac{
p_{{\rm d},k} \,
|\boldsymbol v_k^*\boldsymbol h_{\mathcal M_k, k}|^2
}{
\sum_{m\neq k} \!
\mathbb{E} \big[
|\boldsymbol{v}_k^* \boldsymbol{z}_{k\leftarrow m}[j]|^2 \, |
\boldsymbol{H}
\big]
\! + 
\mathbb E\big[
|\boldsymbol v_k^*\boldsymbol u_k[j]|^2
\, |
\boldsymbol H
\big]
}
\label{eq:sinr_arbitrary_combiner_surrogate_new}
\end{equation}
and the MMSE combiner, $\gamma_k^{\rm MMSE}(\boldsymbol\tau)$, maximizes this SINR over all linear combiners,
A lower bound on the SINR can be obtained by the unit-norm combiner
$\boldsymbol w_k=\frac{\boldsymbol{h}_{\mathcal{M}_k, k}}{\|\boldsymbol{h}_{\mathcal{M}_k, k}\|}$,
which returns the desired signal power 
\begin{align}
p_{{\rm d},k}
|\boldsymbol w_k^*\boldsymbol{h}_{\mathcal{M}_k, k}|^2
&=
p_{{\rm d},k}
\left|
\frac{\boldsymbol{h}_{\mathcal{M}_k, k}^*}{\|\boldsymbol{h}_{\mathcal{M}_k, k}\|}
\boldsymbol{h}_{\mathcal{M}_k, k}
\right|^2 \\&= p_{{\rm d},k}\|\boldsymbol{h}_{\mathcal{M}_k, k}\|^2
\\
&=
p_{{\rm d},k}
\sum_{\ell\in\mathcal M_k}
\|\boldsymbol h_{\ell, k}\|^2.
\label{eq:useful_signal_derived_surrogate_new}
\end{align}
As of the interference contribution from UE $m\neq k$, let
\begin{equation}
\boldsymbol z_{k\leftarrow m}[j]
=
\left[
\boldsymbol z_{\ell_{k,1},k\leftarrow m}[j]^\top,
\ldots,
\boldsymbol z_{\ell_{k,M_k},k\leftarrow m}[j]^\top
\right]^\top.
\label{eq:stacked_interference_surrogate_new}
\end{equation}
Since $\|\boldsymbol w_k\|=1$, 
\begin{equation}
|\boldsymbol w_k^*\boldsymbol z_{k\leftarrow m}[j]|^2
\le
\|\boldsymbol z_{k\leftarrow m}[j]\|^2
\label{eq:cauchy_surrogate_new}
\end{equation}
and thus
\begin{align}
\mathbb E\!\left[
|\boldsymbol w_k^*\boldsymbol z_{k\leftarrow m}[j]|^2
\, |
\boldsymbol H
\right]
& \le
\mathbb E\!\left[
\|\boldsymbol z_{k\leftarrow m}[j]\|^2
\,|
\boldsymbol H
\right]
\\
&=p_{{\rm d},m}
\sum_{\ell\in\mathcal M_k}
\|\boldsymbol h_{\ell, m}\|^2
\Gamma(\tau_k-\tau_m). \notag
\end{align}
Letting $ U_k =
\mathbb E\!\left[
|\boldsymbol w_k^*\boldsymbol u_k[j]|^2
\,|
\boldsymbol H
\right]$,
which includes all the terms not controlled by the pairwise relative data delays,
\begin{equation}
\gamma_k^{\rm MMSE}(\boldsymbol\tau)
\ge
\frac{
p_{{\rm d},k}
\sum_{\ell\in\mathcal M_k}
\|\boldsymbol h_{\ell, k}\|^2
}{
U_k
+
\mathcal J_k(\boldsymbol\tau,\boldsymbol H)
},
\label{eq:MMSE_lower_bound_channel_dependent_surrogate_new}
\end{equation}
where
\begin{equation}
\mathcal J_k(\boldsymbol\tau,\boldsymbol H)
=
\sum_{m\neq k}
p_{{\rm d},m}
\Gamma(\tau_k-\tau_m)
\sum_{\ell\in\mathcal M_k}
\|\boldsymbol h_{\ell, m}\|^2.
\label{eq:Jk_channel_dependent_surrogate_new}
\end{equation}
For fixed large-scale channel conditions and fixed
$U_k$, the lower bound on the MMSE SINR is a
monotonically decreasing function of
$\mathcal J_k(\boldsymbol\tau,\boldsymbol H)$.
With its expected value being
\begin{align}
\overline{\mathcal J}_k(\boldsymbol\tau)
&=
\mathbb E_{\boldsymbol H}
\left[
\mathcal J_k(\boldsymbol\tau,\boldsymbol H)
\right]
\nonumber\\
&=
\sum_{m\neq k}
p_{{\rm d},m}
\Gamma(\tau_k-\tau_m)
\sum_{\ell\in\mathcal M_k}
\operatorname{tr}(\boldsymbol{R}_{\ell, m}),
\label{eq:Jk_large_scale_surrogate_new}
\end{align}
a large-scale delay-assignment surrogate is
\begin{equation}
\min_{\tau_1,\ldots,\tau_K\in\mathcal T}
\sum_{k=1}^{K}
\alpha_k
\overline{\mathcal J}_k(\boldsymbol\tau),
\label{eq:directed_surrogate_new}
\end{equation}
where $\alpha_k\ge 0$ serves to prioritize users.
Combining \eqref{eq:Jk_large_scale_surrogate_new} and
\eqref{eq:directed_surrogate_new} yields
\begin{equation}
\min_{\tau_1,\ldots,\tau_K\in\mathcal T}
\sum_{m<k}
c_{m,k}
\Gamma(\tau_m-\tau_k),
\label{eq:pairwise_surrogate_final_new}
\end{equation}
with
\begin{equation}
c_{m,k}
=
\alpha_m
p_{{\rm d},k}
\sum_{\ell\in\mathcal M_m} \!\!\!
\operatorname{tr}(\boldsymbol{R}_{\ell, k})
+
\alpha_k
p_{{\rm d},m}
\sum_{\ell\in\mathcal M_k} \!\!\!
\operatorname{tr}(\boldsymbol{R}_{\ell, m}).
\label{eq:pairwise_weight_final_new}
\end{equation}
The above minimization coincides with \eqref{eq:bound_minimization_data_delay_problem}, reinforcing its relevance.

\subsection{Spatial-Delay Conflict Graph}
\label{subsec:metrics}


Not all users need different delays. If two users are far apart and do not share serving APs, their relative delay is rather irrelevant. Users $m$ and $k$ 
are relevant in that respect only if they are likely to be hard to separate, say if they are served by common APs, have comparable received strengths on those APs, and are not well separated spatially.

Let us denote heuristic graph weights by $c^{\rm  graph}_{m,k}$, which are different from the derived coefficients $c_{m,k}$. A large value for $c^{\rm graph}_{m,k}$ indicates that Users $m$ and $k$ would benefit from a delay separation, while a small value means that their relative delay is less important. 
Next, heuristic choices are considered for $c^{\rm  graph}_{m,k}$
that import specific features of $c_{m,k}$.

The quantities affecting $c_{m,k}$ are the power gains, the pairwise correlations, and whether the two users are served by the same AP. 
Thus, the general formula for $c_{m,k}^{\rm graph}$ may have the form
\begin{equation}
    c^{\rm graph}_{m,k} = f_1\big(\{\boldsymbol{R}_{\ell, m}\}_{\ell=1}^{L}, \{\boldsymbol{R}_{\ell, k}\}_{\ell=1}^{L}, \{\mathcal A_\ell\}_{\ell=1}^{L}\big).
\end{equation}
A simplified version of this formulation is one where $c^{\rm graph}_{m,k}$ can be represented as a sum of contributions across APs, i.e.,
\begin{equation}
    c^{\rm graph}_{m,k}=\sum_{\ell=1}^L f_2(\boldsymbol{R}_{\ell, m},\boldsymbol{R}_{\ell, k}, \mathcal A_\ell).
\end{equation}
Based on the above, one can start with a conflict metric such as
\begin{equation}
c_{m,k}^{\rm graph}
=
\sum_{\ell=1}^{L}
\boldsymbol{1}(m,k\in\mathcal{A}_\ell) \,
\psi_{\ell, m,k},
\label{eq:conflict_metric}
\end{equation}
where $\boldsymbol{1}(\cdot)$ is the indicator function and $\psi_{\ell,m,k}$ is a measure of the interference coupling between UEs $m$ and $k$ at AP $\ell$.
Several possibilities can be entertained for $\psi_{\ell,m,k}$. 
\begin{itemize}
    \item A simple choice (shared AP-overlap) is
\begin{equation}
c^{\rm graph}_{m,k}
=
\sum_{\ell=1}^{L}
\boldsymbol{1}(m,k\in\mathcal{A}_\ell),
\end{equation}
which counts how many APs serve both users; the premise is that if two users overlap at many APs, they are likely to interfere.
\item An alternative is the large-scale-gain weighted model ($\beta$-product)
\begin{equation}
\psi_{\ell, m,k}=\beta_{\ell, m}\beta_{\ell, k},
\label{eq:beta_conflict}
\end{equation}
where a pair is more relevant if both users are strong at the same AP.
\item Since, if User $m$ is strong and User $k$ is weak, the pair is easily separable, another sensible metric (min-$\beta$) is
\begin{equation}
\psi_{\ell, m,k}=\min(\beta_{\ell, m},\beta_{\ell, k}).
\label{minbeta}
\end{equation}
\item A more intricate metric is
\begin{equation}
\psi_{\ell, m,k}=\beta_{\ell, m}\beta_{\ell, k} \rho_{\ell, m,k},
\end{equation}
where
\begin{equation}
\rho_{\ell,m,k}
=
\frac{
\left|\boldsymbol h_{\ell,m}^{*}\boldsymbol h_{\ell,k}\right|^2
}{
\left\|\boldsymbol h_{\ell,m}\right\|^2
\left\|\boldsymbol h_{\ell,k}\right\|^2
},
\end{equation}
captures the correlation (or alignment) between the channels of User $m$ and User $k$ at AP $\ell$.
\end{itemize}



\section{Solving the Delay Assignment Problem}
\label{sec:delay_assignment}

As a scalable approach to tackle the delay-assignment problem in \eqref{eq:bound_minimization_data_delay_problem}, we posit a greedy weighted-coloring algorithm. First, for each user $m$, define its total conflict weight as
\begin{equation}
    d_m=\sum_{k\ne m} c_{m,k}.
\end{equation}
The users are then ordered in decreasing order of $d_m$, so that the most critical users are assigned first. Next, consider the user $m$ in that order. For each candidate delay $\delta\in\mathcal{T}$, compute the incremental assignment cost
\begin{equation}
    J_m(\delta)=\sum_{k\in\mathcal{S}} c_{m,k}\Gamma(\delta-\tau_k),
\end{equation}
where $\mathcal{S}$ is the set of users whose delays have already been assigned and $\tau_k$ is the delay assigned to User $k$. The delay selected for User $m$ is
\begin{equation}
    \tau_m=\arg\min_{\delta\in\mathcal{T}} J_m(\delta).
\end{equation}
This procedure is repeated until delays are assigned to all users, as summarized in Algorithm~\ref{alg:greedy}.

The formulations expounded in the remainder of this section provide benchmarks for small and moderate problem sizes, and can be used to assess the performance of the greedy algorithm.

\subsection{Binary Delay-Assignment Formulation}

Let us define the binary assignment variables
\begin{equation}
x_{k,q}
=
\begin{cases}
1 & \text{if UE }k\text{ is assigned delay }\delta_q\\
0 & \text{otherwise}.
\end{cases}
\label{eq:x_variable_definition}
\end{equation}
Each UE must be assigned exactly one delay, hence
\begin{equation}
\sum_{q=1}^{Q}x_{k,q}=1
\qquad\quad
k=1,\ldots,K,
\label{eq:one_delay_constraint}
\end{equation}
with $x_{k,q}\in\{0,1\}$.
Precomputing the pulse-overlap table
\begin{equation}
\Gamma_{q,r}
=
\Gamma(\delta_q-\delta_r)
= \!\!
\sum_{u=-U}^{U} \!\!
\left|
q\left((u+\delta_q-\delta_r)T\right)
\right|^2,
\label{eq:gamma_table}
\end{equation}
the objective in \eqref{eq:bound_minimization_data_delay_problem} becomes
\begin{equation}
\min_x
\;
\sum_{m<k}
\sum_{q=1}^{Q}
\sum_{r=1}^{Q}
c_{m,k}\Gamma_{q,r}x_{m,q}x_{k,r}.
\label{eq:bqp_delay_assignment}
\end{equation}
Together with $x_{k,q}\in\{0,1\}$ and \eqref{eq:one_delay_constraint}, this is a binary quadratic program.

\subsection{MILP Formulation}

The quadratic products in \eqref{eq:bqp_delay_assignment} can be linearized
exactly. Let us introduce the auxiliary variables
\begin{equation}
y_{m,k,q,r}=x_{m,q}x_{k,r}
\qquad
m<k,\quad q,r=1,\ldots,Q.
\label{eq:y_product_definition}
\end{equation}
As $x_{m,q}$ and $x_{k,r}$ are binary, the equality
$y_{m,k,q,r}=x_{m,q}x_{k,r}$ is enforced by the linear constraints
\begin{align}
y_{m,k,q,r} &\le x_{m,q}
\label{eq:mccormick_1}\\
y_{m,k,q,r} &\le x_{k,r}
\label{eq:mccormick_2}\\
y_{m,k,q,r} &\ge x_{m,q}+x_{k,r}-1
\label{eq:mccormick_3}\\
y_{m,k,q,r} &\ge 0.
\label{eq:mccormick_4}
\end{align}
The delay-assignment problem can be written as the mixed-integer linear program (MILP)
\label{eq:milp_delay_assignment}
\begin{align}
\min_{x,y}
\quad
&
\sum_{m<k}
\sum_{q=1}^{Q}
\sum_{r=1}^{Q}
c_{m,k}\Gamma_{q,r}y_{m,k,q,r}
\label{eq:milp_obj}
\\
\mathrm{s.t.}
\quad
&
\sum_{q=1}^{Q}x_{k,q}=1
&& \!\!\! k=1,\ldots,K,
\notag 
\\
&
y_{m,k,q,r}\le x_{m,q}
&& \!\!\! m<k,\; q,r=1,\ldots,Q
\notag 
\\
&
y_{m,k,q,r}\le x_{k,r}
&& \!\!\! m<k,\; q,r=1,\ldots,Q
\notag 
\\
&
y_{m,k,q,r}\ge x_{m,q}+x_{k,r}-1
&& \!\!\! m<k,\; q,r=1,\ldots,Q
\notag 
\\
&
x_{k,q}\in\{0,1\}
&& \!\!\! k=1,\ldots,K \notag \\
& && \!\!\! q=1,\ldots,Q
\notag 
\\
&
y_{m,k,q,r}\ge 0
&& \!\!\! m<k,\; q,r=1,\ldots,Q,
\notag 
\end{align}
to which standard solvers can be applied.
The MILP-based procedure is detailed in Algorithm~\ref{alg:interference}.

\begin{algorithm}[t]
\caption{Greedy weighted-coloring delay assignment}
\label{alg:greedy}
\begin{algorithmic}[1]
\Require Delay codebook, $\mathcal T=\{\delta_1,\ldots,\delta_Q\}$; pairwise weights, $\{c_{m,k}\}$; pulse-overlap function, $\Gamma(\cdot)$
\Ensure Delay assignment $\boldsymbol{\tau}$
\State Compute $d_m=\sum_{k\ne m}c_{m,k}$ for $m=1,\ldots,K$
\State Sort the users in decreasing order of $d_m$
\State $\mathcal{S}\gets\emptyset$
\For{each user $m$ in that order}
    \For{each $\delta\in\mathcal{T}$}
        \State $J_m(\delta)=\sum_{k\in\mathcal{S}}c_{m,k}\Gamma(\delta-\tau_k)$
    \EndFor
    \State $\tau_m\gets\argmin_{\delta\in\mathcal{T}}J_m(\delta)$
    \State $\mathcal{S}\gets\mathcal{S}\cup\{m\}$
\EndFor
\State \Return $\boldsymbol{\tau}$
\end{algorithmic}
\end{algorithm}

\begin{algorithm}[t]
\caption{MILP-based delay assignment 
}
\label{alg:interference}
\begin{algorithmic}[1]
\Require Delay codebook, $\mathcal T=\{\delta_1,\ldots,\delta_Q\}$; large-scale gains, $\{\beta_{\ell, k}\}$; data powers, $\{p_{{\rm d},k}\}$; serving sets, $\{\mathcal M_k\}$; pulse autocorrelation, $q(t)$; pulse memory, $U$
\Ensure Delay assignment $\widehat{\boldsymbol{\tau}}$

\Statex
\State \textbf{Compute pairwise interference weights}
\For{$m=1,\ldots,K-1$}
    \For{$k=m+1,\ldots,K$}
        \State 
        $c_{m,k} = 
        p_{{\rm d},k}\sum_{\ell\in\mathcal M_m} \! \operatorname{tr}(\boldsymbol{R}_{\ell, k})$ \\
 \qquad \qquad\qquad   $ + \;  p_{{\rm d},m}\sum_{\ell\in\mathcal M_k} \! \operatorname{tr}(\boldsymbol{R}_{\ell, m})$
    \EndFor
\EndFor

\Statex
\State \textbf{Compute pulse-overlap table}
\For{$q=1,\ldots,Q$}
    \For{$r=1,\ldots,Q$}
        \State Compute
        $\Gamma_{q,r}
        =
        \sum_{u=-U}^{U}
        \left|
        q\left((u+\delta_q-\delta_r)T\right)
        \right|^2$.
    \EndFor
\EndFor

\Statex
\State \textbf{Build MILP variables}
\State Introduce binary variables $x_{k,q}\in\{0,1\}$, where $x_{k,q}=1$ if UE $k$ is assigned delay $\delta_q$
\State Introduce auxiliary variables $y_{m,k,q,r}\ge 0$ to represent the products $x_{m,q}x_{k,r}$

\Statex
\State \textbf{Solve the MILP (Sec. \ref{eq:milp_delay_assignment})}

\State \textbf{Recover assigned delays}
\For{$k=1,\ldots,K$}
    \State Set $\widehat{\tau}_k=\delta_q$ where $x_{k,q}=1$
\EndFor

\State \Return $\widehat{\boldsymbol{\tau}}$
\end{algorithmic}
\end{algorithm}

\section{Performance Evaluation}
\label{sec:simulation_results}

This section
evaluates the performance improvement associated with the assignment of delays, while also verifying the impact of receiver constraints such as windowed MMSE processing, the finite delay-codebook size, AP load limits, channel estimation, and local processing.

APs and UEs are independently and uniformly dropped in a square wrapped-around deployment area. For each drop, the spectral efficiency is averaged over small-scale Rayleigh fading realizations.
The AP association is user-centric: each UE is served by its strongest APs subject to a per-AP maximum load constraint.
The asynchronous delays are chosen from a finite uniform delay codebook (see Sec. \ref{subsec:codebook}),
with Table \ref{tab:simulation_parameters} detailing all the simulation parameters.
The large-scale gains follow \eqref{eq:channel_power} with $\beta_0=-30.5$~dB and $\kappa=3.67$, where $D_{\ell,k}$ is the three-dimensional AP-UE distance, multiplied by a log-normal shadowing factor.
Spectral efficiency includes the prelog factor $\frac{N_\text{d}}{N_\text{d}+N_\text{p}+1}$, where the added 1 in the denominator accounts for the gap between synchronous pilots and asynchronous data symbols.

\begin{table}[t]
\centering
\caption{Simulation parameters.}
\label{tab:simulation_parameters}
\renewcommand{\arraystretch}{1.22}
\begin{tabular}{|l|l|}
\hline
\textbf{Parameter} & \textbf{Value} \\
\hline
Number of APs, $L$ & $20$ \\
\hline
Number of UEs, $K$ & $16$ \\
\hline
Antennas per AP, $N_{\text R}$ & $1$ \\
\hline
Serving APs per UE, $M_{\rm max}$ & $4$ strongest \\
\hline
Maximum served UEs per AP, $S_{\max}$ & $8$ \\
\hline
Deployment area & $1000 \times 1000$ m$^2$ \\
\hline
Bandwidth, $B$ (MHz) & $20$ \\
\hline
Symbol period, $T$ & $1/B$ \\
\hline
Data symbols length, $N_\text{d}$ & $128$ \\
\hline
Pilot length, $N_\text{p}$ &  $K$ \\
\hline
Uplink data power, $p_\text{d}$ (W) & $0.2$ \\
\hline
Pilot power, $p_\text{p}$ (W) & $0.2$ \\
\hline
Delay-codebook size, $Q$ & $4$ \\
\hline
Default MMSE window half-width, $W$ & $3$ symbols \\
\hline
Delay-objective truncation, $U$ & $2$ \\
\hline
Pulse shape & RRC ($\upbeta =1$) or rectangular\\
\hline
Noise figure (dB) & $7$ \\
\hline
Pathloss (dB) & $-30.5-36.7\log_{10} D_{lk}$\\ &+ shadowing \\
\hline
Shadow fading standard deviation (dB) & $4$ \\
\hline
Shadow decorrelation distance (m) & $9$ \\
\hline
AP-UE vertical distance (m) & $10$ \\
\hline
Monte-Carlo drops & $200$ \\
\hline
Small-scale realizations per drop & $100$ \\
\hline
\end{tabular}
\end{table}

\begin{figure}[t]
\centering
\includegraphics[width=\linewidth]{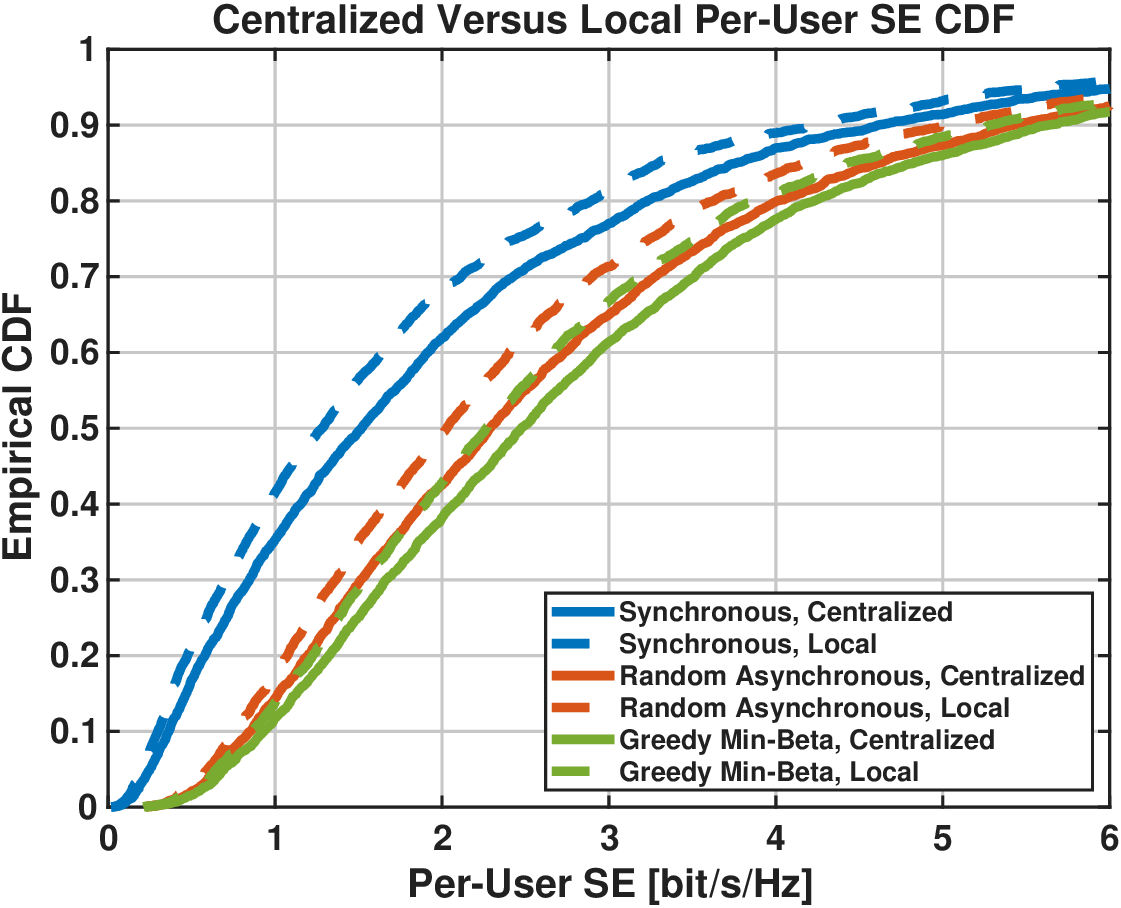}
\caption{CDF of ergodic per-user spectral efficiency for centralized and local
processing with rectangular pulses.}
\label{fig:central_vs_local_rate_cdf}
\end{figure}

\begin{figure}[t]
\centering
\includegraphics[width=\linewidth]{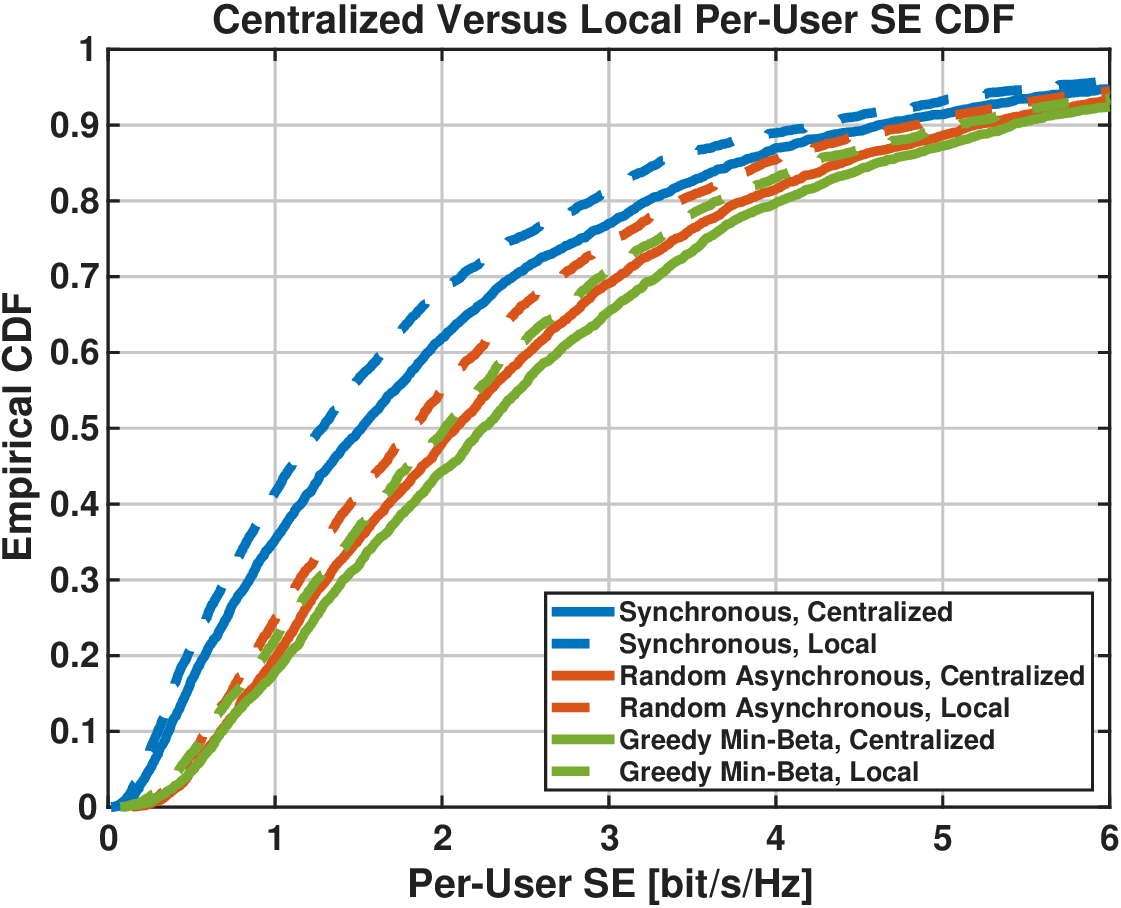}
\caption{CDF of ergodic per-user spectral efficiency for centralized and local
processing with RRC pulses.}
\label{fig:rrc_central_vs_local_rate_cdf}
\end{figure}

\begin{figure}[t]
\centering
\includegraphics[width=\linewidth]{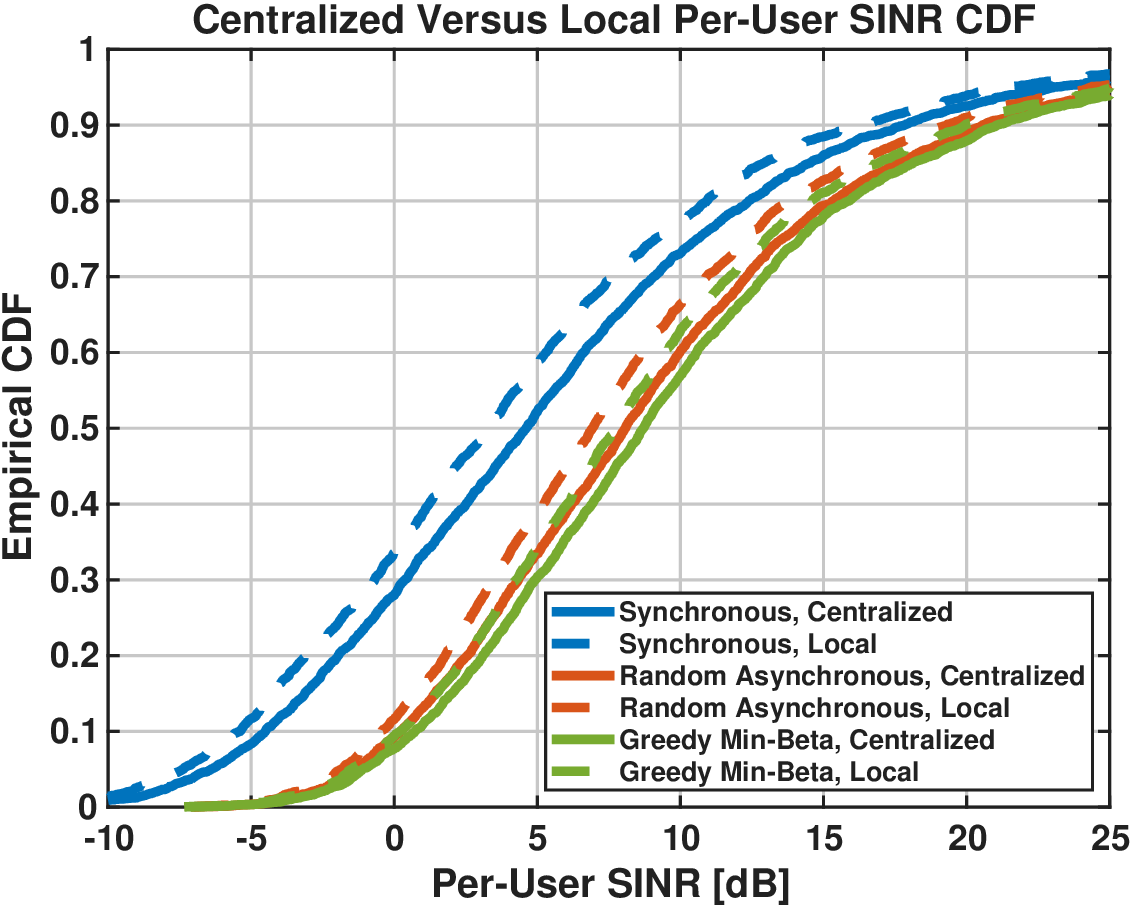}
\caption{CDF of local-average SINR for centralized and local processing with rectangular pulses.}
\label{fig:central_vs_local_sinr_cdf}
\end{figure}

\begin{figure}[t]
\centering
\includegraphics[width=\linewidth]{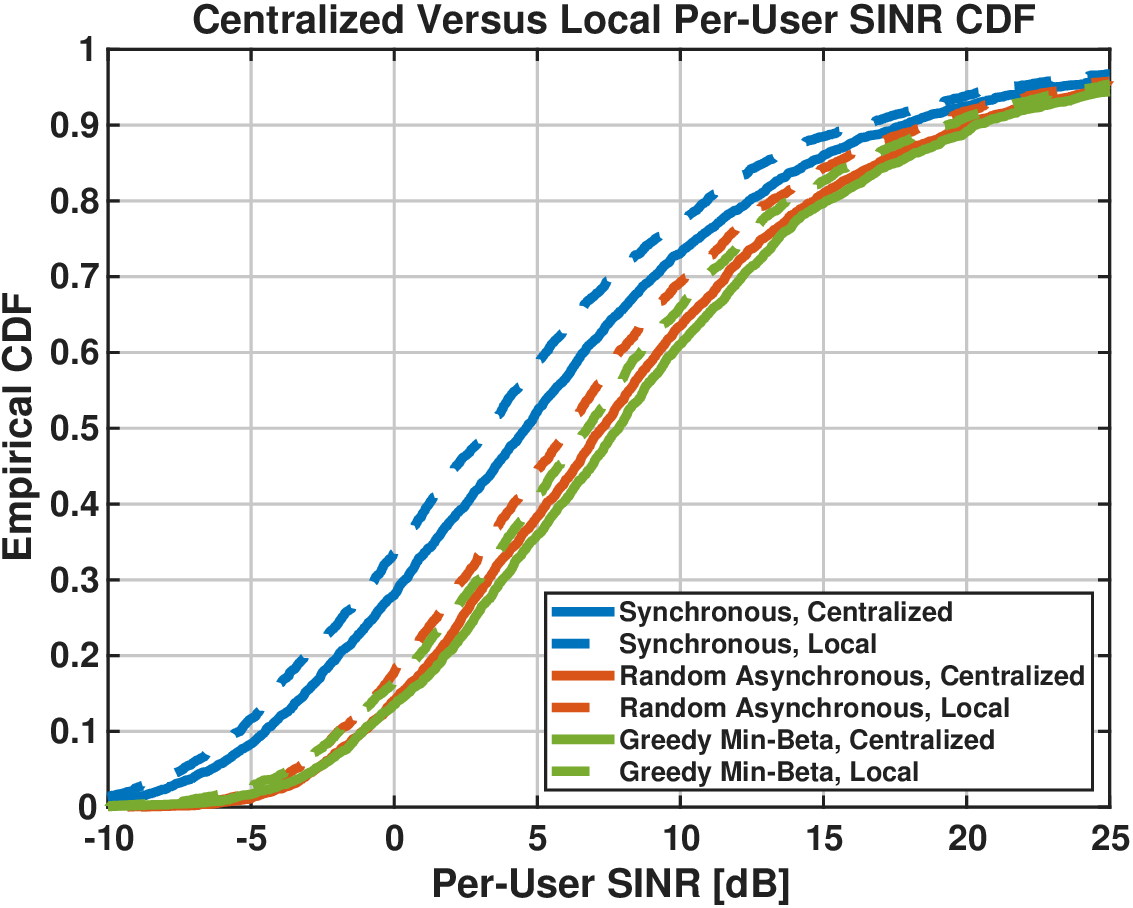}
\caption{CDF of local-average SINR for centralized and local
processing with RRC pulses.}
\label{fig:rrc_central_vs_local_sinr_cdf}
\end{figure}

To establish the potential of introducing controlled delays,
Fig.~\ref{fig:central_vs_local_rate_cdf} compares the performance of (i) a synchronous network, (ii) an asynchronous network with random delays, and (iii) an asynchronous network with delays set by the greedy weight coloring algorithm with the graph weights in \eqref{minbeta}, based on the minimum large-scale gain $\beta$.
The comparison is presented for both
centralized and local processing, by means of the cumulative distribution function (CDF) of
per-user spectral efficiency with rectangular pulses. The comparison is repeated, for root raised cosine (RRC) pulses as an embodiment of Nyquist pulses, in Fig.~\ref{fig:rrc_central_vs_local_rate_cdf}.
Then, Figs.~\ref{fig:central_vs_local_sinr_cdf} and~\ref{fig:rrc_central_vs_local_sinr_cdf} show the respective SINR distributions.
In all cases, time-invariant combiners are employed; precisely, the combiner computed for the middle symbol in every frame is applied over that entire frame.\footnote{A time-variant combiner would further increase the spectral efficiency, but very slightly---less than $1\%$ for the parameters in Table \ref{tab:simulation_parameters} over the operating regimes of interest.}
The results show that the introduction of delays markedly improves the performance, with the specific greedy min-$\beta$ strategy further extending the advantage relative to random delays. Expectedly, centralized processing has a performance edge over local processing, at the cost of higher  complexity.

Elaborating on the greedy min-$\beta$ approach, Table~\ref{tab:rectangular_codebook_size} quantifies its performance as a function of the number of delay codewords, $Q$. A larger value gives the algorithm more freedom to separate strongly interfering users, but it increases the number of delay branches that APs may need to sample. The welcome observation is that small values suffice, with rapidly diminishing returns thereafter. 





\begin{figure}[t]
\centering
\includegraphics[width=\linewidth]{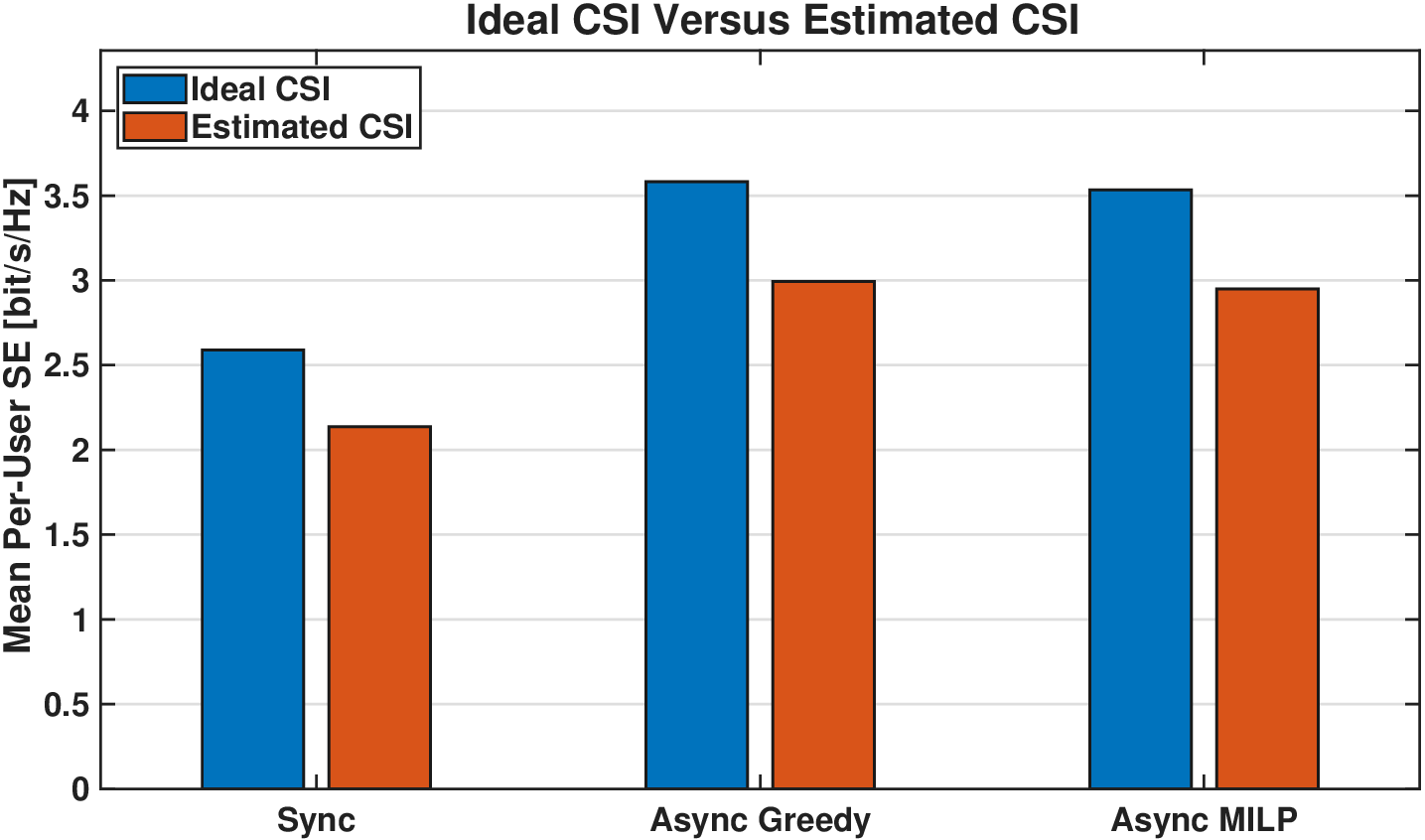}
\caption{Average per-user spectral efficiency under full CSI and synchronous-pilot-based estimated CSI with rectangular pulses.}
\label{fig:full_vs_estimated}
\end{figure}

To assess the impact of channel estimation,
Fig.~\ref{fig:full_vs_estimated} compares the average ergodic spectral efficiency throughout the network under ideal channel state information (CSI) and under estimated CSI for the synchronous and asynchronous schemes. Channel estimation does incur a penalty, largely due to the pilot overhead, but the benefits of asynchronicity are not diminished.
Also relevant is that the proposed greedy algorithm is highly effective, essentially matching the performance bound embodied by the MILP.


\begin{table}[t]
\centering
\caption{Impact of delay-codebook size on centralized time-varying MMSE processing with rectangular pulses.}
\label{tab:rectangular_codebook_size}
\renewcommand{\arraystretch}{1.2}
\begin{tabular}{|c|c|c|c|}
\hline
$Q$ &
\shortstack{Mean SE\\{}[bit/s/Hz/user]} &
\shortstack{10\%-likely SE\\{}[bit/s/Hz/user]} &
\shortstack{Median SINR\\{}[dB]} \\
\hline
1 & 2.088 & 0.362 & 2.96 \\
2 & 2.775 & 0.689 & 6.67 \\
4 & 3.002 & 0.950 & 7.68 \\
6 & 3.030 & 0.966 & 7.77 \\
8 & 3.037 & 0.992 & 7.81 \\
\hline
\end{tabular}
\end{table}


Tables~\ref{tab:ap_cap_rectangular} and \ref{tab:ap_cap_rrc} illustrate how the performance depends on $S_{\max}$, the maximum number of users served by each AP for rectangular and RRC pulses, respectively. A small number of users $S_{\max}=4$ suffices for the performance to settle.



\begin{table}[t]
\centering
\caption{Impact of $S_{\max}$ with rectangular pulses.}
\label{tab:ap_cap_rectangular}
\renewcommand{\arraystretch}{1.2}
\begin{tabular}{|c|c|c|c|}
\hline
$S_{\max}$ &
\shortstack{Mean SE\\{}[bit/s/Hz]} &
\shortstack{10\%-likely SE\\{}[bit/s/Hz]} &
\shortstack{Median SINR\\{}[dB]} \\
\hline
4  & 2.953 & 0.903 & 7.50 \\
6  & 2.973 & 0.944 & 7.58 \\
8  & 2.974 & 0.935 & 7.60 \\
10 & 2.974 & 0.935 & 7.60 \\
\hline
\end{tabular}
\end{table}


\begin{table}[t]
\centering
\caption{Impact of $S_{\max}$ with RRC pulses ($\beta=1$).}
\label{tab:ap_cap_rrc}
\renewcommand{\arraystretch}{1.2}
\begin{tabular}{|c|c|c|c|}
\hline
$S_{\max}$ &
\shortstack{Mean SE\\{}[bit/s/Hz]} &
\shortstack{10\%-likely SE\\{}[bit/s/Hz]} &
\shortstack{Median SINR\\{}[dB]} \\
\hline
4  & 2.724 & 0.669 & 6.47 \\
6  & 2.747 & 0.692 & 6.59 \\
8  & 2.748 & 0.691 & 6.59 \\
10 & 2.748 & 0.691 & 6.59 \\
\hline
\end{tabular}
\end{table}


To assess the impact of antenna correlations within the APs, we next let $N_\text{R}=2$.
In Table~\ref{tab:correlated_vs_uncorrelated}, various delay assignment schemes are evaluated without correlation and with a correlation factor of $0.7$ 
between the antennas.
The considered schemes are based on the greedy procedure in Algorithm~\ref{alg:interference}, with pairwise conflict weights set forth in Sec.~\ref{subsec:metrics}.
Across the board, antenna correlation produces a modest reduction in the absolute performance of all schemes, but the benefit of intentional asynchronicity remains essentially unchanged.

\begin{table}[t]
\centering
\caption{Two-antenna rectangular-pulse performance with uncorrelated and correlated ($\rho=0.7$) fading.}
\label{tab:correlated_vs_uncorrelated}
\resizebox{\columnwidth}{!}{%
\begin{tabular}{|l|cc|cc|}
\hline
Scheme & \multicolumn{2}{c|}{Mean SE [bit/s/Hz/user]} & \multicolumn{2}{c|}{Median SINR [dB]} \\
& Uncorrelated & Correlated & Uncorrelated & Correlated \\
\hline
Synchronous & 2.985 & 2.854 & 8.08 & 7.55 \\
Random delays & 3.713 & 3.554 & 11.11 & 10.54 \\
Greedy $\beta$-product & 3.779 & 3.627 & 11.21 & 10.63 \\
Greedy $\beta$-product $\times$ alignment & 3.795 & 3.637 & 11.29 & 10.69 \\
Greedy shared-AP overlap & 3.855 & 3.708 & 11.55 & 11.04 \\
Greedy min-$\beta$ & 3.863 & 3.718 & 11.54 & 11.03 \\
Greedy min-$\beta$ $\times$ alignment & 3.873 & 3.727 & 11.64 & 11.10 \\
\hline
\end{tabular}%
}
\end{table}

\section{Conclusion}
\label{sec:conclusion}

This paper has studied intentional symbol-level asynchrony as a design resource for uplink user-centric cell-free networks.
It has been shown that purposely assigned delays can reduce pulse overlap among interfering users, thereby improving the performance. It has further been shown that finite-delay codebooks, delay-aware assignment rules, and bounded per-AP sampling keep the receiver complexity manageable while retaining most of the performance benefit of asynchronous reception. Results indicate that
simple greedy delay assignment schemes perform quasi-optimally.
Future work may extend the framework to joint delay and AP-cluster optimization, more general pulse shapes, imperfect synchronization constraints, and scalable receiver implementations that further reduce the cost of performance evaluation in large networks.

\appendices
\section{} 
\label{app:proof_disturbance_cov}

    Substituting $\boldsymbol{\eta}_{\ell,k}$ from \eqref{eq:eta_lk} into \eqref{eq:disturbance_cov},
\begin{align}\label{eq:covariance_disturbance}
    \boldsymbol{C}_{\boldsymbol{\eta}_{\ell, k}}
    &=
    \frac{1}{N_\text{p}^2}
    \sum_{m=1}^{K}
    p_{{\rm d},m}
    \mathbb{E}\left[
    \left|
    \bigl(\boldsymbol{\mathsf{A}}_{m}\boldsymbol{s}_m\bigr)^\top
    \bar{\boldsymbol{\phi}}_k
    \right|^2
    \right]
    \boldsymbol{R}_{\ell, m}
    \nonumber +\boldsymbol{\mathsf{C}}_{\boldsymbol{\mathsf{w}_{\ell, k}}}.
\end{align}
The scalar term inside the expectations can be rewritten as
\begin{align}
    \mathbb{E} \! \left[
    \left|
    \bigl(\boldsymbol{\mathsf{A}}_{m}\boldsymbol{s}_m\bigr)^\top
    \bar{\boldsymbol{\phi}}_k
    \right|^2
    \right]
    =
    \boldsymbol{\phi}_k^\top
    \bar{\boldsymbol{\mathsf{A}}}_{m}
    \boldsymbol{\mathsf{A}}_{m}^\top
    \bar{\boldsymbol{\phi}}_k.
\end{align}
As the pilot sequences are orthogonal and $p(t)$ is unit-energy, the filtered noise covariance satisfies
$
    \boldsymbol{\mathsf{C}}_{\boldsymbol{\mathsf{w}_{\ell, k}}}
    =
    (\sigma^2/N_\text{p}) \boldsymbol{I}_{N_\text{R}}.
$

\section{}
\label{app:proof_residual_cov}
    Define
\begin{align}
    \boldsymbol{u}_{m,k}
    =
    \boldsymbol{\mathsf{B}}_{k}\boldsymbol{\phi}_m
    \in \mathbb{C}^{N_\text{d}\times 1}.
\end{align}
Using the identity
\begin{align}
    (\boldsymbol{I}_{N_\text{d}}\otimes \boldsymbol{x})\boldsymbol{u}
    =
    \boldsymbol{u}\otimes \boldsymbol{x},
\end{align}
the residual pilot-interference term can be rewritten as
\begin{align}
    \boldsymbol{\mathsf{r}}^{{\rm res}}_{\ell, k}
    =
    \sum_{m=1}^{K}
    \sqrt{p_{{\rm p},m}}
    \left(
    \boldsymbol{u}_{m,k}\otimes \tilde{\boldsymbol{h}}_{\ell, m}
    \right).
\end{align}
Similarly,
\begin{align}
    \boldsymbol{\mathsf{r}}^{{\rm res}}_{\ell, k'}
    =
    \sum_{n=1}^{K}
    \sqrt{p_{{\rm p},n}}
    \left(
    \boldsymbol{u}_{n,k'}\otimes \tilde{\boldsymbol{h}}_{\ell, n}
    \right).
\end{align}
Therefore,
\begin{align}
    &\boldsymbol{\mathsf{C}}^{{\rm res}}_{\ell,k,k'}
    =
    \mathbb{E} \! \left[
    \boldsymbol{\mathsf{r}}^{{\rm res}}_{\ell, k}
    \bigl(\boldsymbol{\mathsf{r}}^{{\rm res}}_{\ell, k'}\bigr)^{*}
    \right]
    \\
    &=
    \sum_{m=1}^{K}
    \sum_{n=1}^{K}
    \sqrt{p_{{\rm p},m}p_{{\rm p},n}}
    \mathbb{E}\left[
    \left(
    \boldsymbol{u}_{m,k}\otimes \tilde{\boldsymbol{h}}_{\ell, m}
    \right)
    \left(
    \boldsymbol{u}_{n,k'}\otimes \tilde{\boldsymbol{h}}_{\ell, n}
    \right)^{*}
    \right]. \notag
\end{align}
Using the Kronecker identity
\begin{align}
    (\boldsymbol{a}\otimes \boldsymbol{x})
    (\boldsymbol{b}\otimes \boldsymbol{y})^{*}
    =
    (\boldsymbol{a}\boldsymbol{b}^{*})
    \otimes
    (\boldsymbol{x}\boldsymbol{y}^{*}),
\end{align}
we obtain
\begin{align}
    \boldsymbol{\mathsf{C}}^{{\rm res}}_{\ell,k,k'}
    &=
    \sum_{m=1}^{K}
    \sum_{n=1}^{K}
    \sqrt{p_{{\rm p},m}p_{{\rm p},n}}
    \left(
    \boldsymbol{u}_{m,k}\boldsymbol{u}_{n,k'}^{*}
    \right)
    \otimes
    \mathbb{E} \! \left[
    \tilde{\boldsymbol{h}}_{\ell, m}
    \tilde{\boldsymbol{h}}_{\ell, n}^{*}
    \right]. \notag
\end{align}
Since the channel-estimation errors of different UEs are uncorrelated,
\begin{align}
    \mathbb{E} \! \left[
    \tilde{\boldsymbol{h}}_{\ell, m}
    \tilde{\boldsymbol{h}}_{\ell, n}^{*}
    \right]
    =
    \begin{cases}
        \boldsymbol{C}_{\tilde{\boldsymbol{h}}_{\ell, m}} & \quad m=n\\
        \boldsymbol{0} & \quad m\neq n.
    \end{cases}
\end{align}
Hence, all terms with $m\neq n$ vanish,
\begin{align}
    \boldsymbol{\mathsf{C}}^{{\rm res}}_{\ell,k,k'}
    &=
    \sum_{m=1}^{K}
    p_{{\rm p},m}
    \left(
    \boldsymbol{u}_{m,k}\boldsymbol{u}_{m,k'}^{*}
    \right)
    \otimes
    \boldsymbol{C}_{\tilde{\boldsymbol{h}}_{\ell, m}}.
\end{align}
Substituting
$\boldsymbol{u}_{m,k}
=
\boldsymbol{\mathsf{B}}_{k}\boldsymbol{\phi}_m$
and
$\boldsymbol{u}_{m,k'}
=
\boldsymbol{\mathsf{B}}_{k'}\boldsymbol{\phi}_m$,
we obtain
\begin{align}
    \boldsymbol{\mathsf{C}}^{{\rm res}}_{\ell,k,k'}
    &=
    \sum_{m=1}^{K}
    p_{{\rm p},m}
    \left(
    \boldsymbol{\mathsf{B}}_{k}\boldsymbol{\phi}_m
    \boldsymbol{\phi}_m^{*}
    \boldsymbol{\mathsf{B}}_{k'}^{*}
    \right)
    \otimes
    \boldsymbol{C}_{\tilde{\boldsymbol{h}}_{\ell, m}}.
\end{align}

\bibliography{journalAbbreviations, references2}
\bibliographystyle{ieeetr}
\end{document}